\documentclass[9pt]{ieeetj_rev}
\usepackage{cite}
\usepackage{amsmath,amssymb,amsfonts}
\usepackage{algorithmic}
\usepackage{graphicx,color}
\usepackage{textcomp}
\usepackage{xcolor}
\usepackage{hyperref}
\hypersetup{hidelinks}
\usepackage{balance}
\usepackage{algorithm,algorithmic}
\def\BibTeX{{\rm B\kern-.05em{\sc i\kern-.025em b}\kern-.08em
    T\kern-.1667em\lower.7ex\hbox{E}\kern-.125emX}}
\AtBeginDocument{\definecolor{tmlcncolor}{cmyk}{0.93,0.59,0.15,0.02}\definecolor{NavyBlue}{RGB}{0,86,125}}

\usepackage{algorithm}
\usepackage{array}
\usepackage{stfloats}
\usepackage{url}
\usepackage{verbatim}
\usepackage{cite}
\usepackage{subfigure}
\usepackage{caption, subcaption, amsmath, graphicx, amssymb, changepage}
\usepackage{microtype}
\usepackage{mathtools}
\usepackage{amsthm}
\usepackage{siunitx}
\usepackage{booktabs}
\let\labelindent\relax %
\usepackage{enumitem}
\usepackage{multirow}
\usepackage{arydshln}
\makeatletter
\def\adl@drawiv#1#2#3{%
        \hskip.5\tabcolsep
        \xleaders#3{#2.5\@tempdimb #1{1}#2.5\@tempdimb}%
                #2\z@ plus1fil minus1fil\relax
        \hskip.5\tabcolsep}
\newcommand{\cdashlinelr}[1]{%
  \noalign{\vskip\aboverulesep
           \global\let\@dashdrawstore\adl@draw
           \global\let\adl@draw\adl@drawiv}
  \cdashline{#1}
  \noalign{\global\let\adl@draw\@dashdrawstore
           \vskip\belowrulesep}
           }
\makeatother

\newcommand{\acc}{\text{acc}}
\newcommand{\Att}{\text{Att}}
\newcommand{\med}{\text{med}}
\newcommand{\std}{\text{std}}
\newcommand{\sigmoid}{\text{sigmoid}}
\newcommand{\cC}{\mathcal{C}}
\newcommand{\cA}{\mathcal{A}}
\newcommand{\cH}{\mathcal{H}}

\def\OJlogo{\vspace{-4pt} }
\def\seclogo{\vspace{10pt} }

\def\authorrefmark#1{\ensuremath{^{\textbf{#1}}}}

\begin{document}

\markboth{}{Koo {et al.}}

\title{SMITIN: Self-Monitored Inference-Time INtervention for Generative Music Transformers}

\author{Junghyun Koo\authorrefmark{1, 2}, Member, IEEE, 
Gordon Wichern\authorrefmark{1}, Member, IEEE, \\
François G.\ Germain\authorrefmark{1}, Member, IEEE, Sameer Khurana\authorrefmark{1}, Member, IEEE,\\ and Jonathan Le Roux\authorrefmark{1}, Fellow, IEEE}
\affil{Mitsubishi Electric Research Laboratories (MERL), Cambridge, MA 02139 USA}
\affil{Seoul National University, Seoul, South Korea}
\authornote{This work was performed while J.~Koo was an intern at MERL.}

\begin{abstract}
We introduce Self-Monitored Inference-Time INtervention (SMITIN), an approach for controlling an autoregressive generative music transformer using classifier probes. These simple logistic regression probes are trained on the output of each attention head in the transformer using a small dataset of audio examples both exhibiting and missing a specific musical trait (e.g., the presence/absence of drums, or real/synthetic music).  We then steer the attention heads in the probe direction, ensuring the generative model output captures the desired musical trait. Additionally, we monitor the probe output to avoid adding an excessive amount of intervention into the autoregressive generation, which could lead to temporally incoherent music. We validate our results objectively and subjectively for both audio continuation and text-to-music applications, demonstrating the ability to add controls to large generative models for which retraining or even fine-tuning is impractical for most musicians.

Audio samples of the proposed intervention approach are available on our \underline{\href{http://tinyurl.com/smitin}{demo page}}.
\end{abstract}


\maketitle

\section{Introduction}
\label{sec:intro}

Dynamic control of an audio signal can be considered a fundamental job of most audio professionals. Musicians must precisely control their instruments, and recording studio engineers manipulate various controls to achieve a desired result. Recent advances in the generative modeling of text and images have also been applied to audio signals, leading to an emerging body of literature on how to condition and control audio generative models \cite{jukebox, musicgen, audioldm2, wu2023musicControlNet, garcia2023vampnet, kreuk2022audiogen, borsos2023audiolm}.

\begin{figure}[t]
     \centering
     \includegraphics[width=0.45\textwidth]{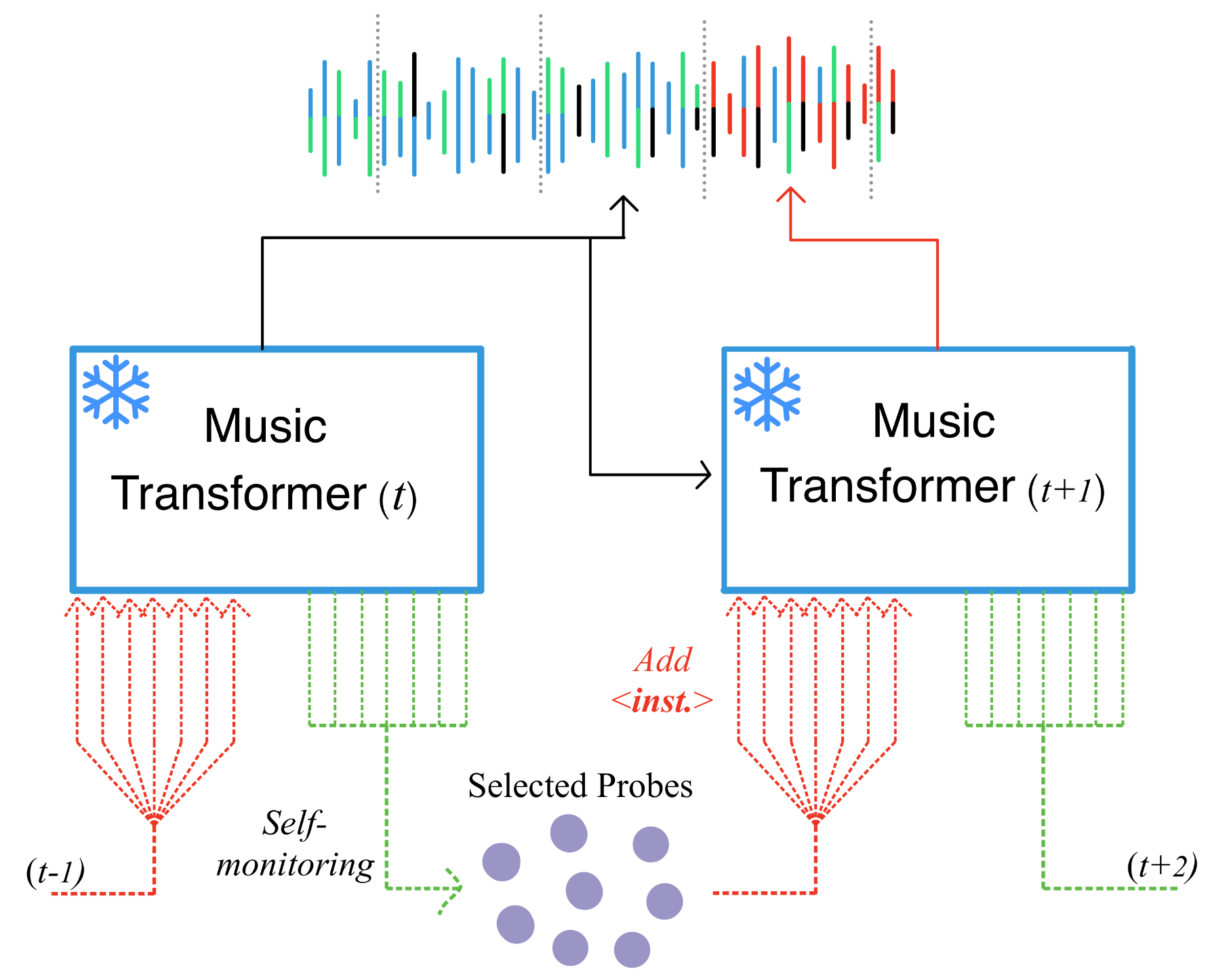}
     \caption{Overall pipeline of SMITIN for inference-time intervention on a pre-trained music generative transformer. The process attempts to enforce specific musical factors (e.g., presence of a particular instrument) during the generation process. SMITIN utilizes a self-monitoring technique to dynamically adjust the intervention strength at each generation step, enabling precise control over the inclusion of the target characteristic while preserving the musical integrity of the output.}
    \label{fig:overall}
    \vspace{-.6cm} %
\end{figure}

Among possible conditioning inputs, describing the audio a user wishes to generate in terms of natural language arguably provides the most flexibility. Such text-to-music models, however, lack the fine-grained temporal control often desired by audio professionals. This has led to training text-to-music models that allow additional conditioning inputs, such as melody sequences \cite{musicgen}, or other types of sequential input \cite{hawthorne2022multi, donahue2023singsong, parker2024stemgen, nistal2024diff}, but these models still suffer from the fact that, once trained, the type of conditioning input is fixed. To alleviate this limitation, some works have explored the addition of new types of control inputs without requiring retraining~\cite{levy2023controllable, wu2023musicControlNet, novack2024ditto, lin2023content, lin2024arrange, zhang2024musicmagus}.  The standard approach in most previous work trains supplementary adapters (e.g., LoRA~\cite{Hu2021}) atop the foundational generative model to introduce additional control, but once the adaptation is complete, it is often not straightforward to vary the strength of the control at inference time. Our goal is to build  ``knob-like’’ variable strength controls given only a small number of audio examples both with and without a desired musical trait (i.e., binary labels).

Audio ``language models'' typically work by first tokenizing chunks of an audio waveform, for example by passing them through the encoder of a pre-trained residual vector quantized autoencoder~\cite{zeghidour2021soundstream, defossez2022high, kumar2023high}, and then pass these audio tokens through an autoregressive transformer trained to predict the next token. Finally, a decoder converts the generated token back into audio. While not state-of-the-art in terms of generation quality compared to latent diffusion~\cite{evans2024stable, mariani2023multi} or masked token models~\cite{garcia2023vampnet, ziv2024masked}, autoregressive audio generative models are still useful for many real-time and interactive applications.  Furthermore, because this class of text-to-music models shares many architectural characteristics with large language models (LLMs), it is interesting to explore whether techniques developed to provide inference-time control of LLMs~\cite{subramani2022extracting, turner2023activation, hernandez2023inspecting} may also be effective for audio.  In particular, we take inspiration from recent work attempting to make text language models more truthful by taking advantage of the fact that the learned internal representations of these models are able to represent the concept of truthfulness~\cite{iti}. Classifier probes~\cite{alain2016understanding, belinkov2022probing, tenney2019bert} applied to each attention head are used to determine whether or not that attention head has learned to represent truthfulness, and inference-time intervention (ITI) is then used to surgically modify the outputs of only the most truth-correlated attention heads, resulting in an LLM that hallucinates less. Previous work~\cite{jukebox_probing} has shown that classifier probes trained on the internal representations of generative music transformers can lead to strong performance across a wide variety of music classification tasks. However, these probes were trained using the learned representation from entire layers, not individual attention heads.

We first show using classifier probes that individual self-attention heads in a pre-trained autoregressive music transformer, MusicGen~\cite{musicgen}, have indeed learned to represent aspects of music we may wish to build control knobs for, making it a priori suitable for ITI approaches such as in~\cite{iti}. However, we find that direct application of this particular approach to a music transformer is sub-optimal because of the difficulty in generating long temporally coherent music samples, as too much intervention causes the generated music to quickly become incoherent. Furthermore, the method asks for the empirical tuning of the number of heads selected for intervention, impeding the ability to implement multiple intervention kinds at scale. To remedy this, we propose the inclusion of a self-monitoring process into the intervention operation, such that we only apply the intervention when the learned classifier probes tell us it is necessary based on the state of the generation network.
Crucially, this self-monitoring technique enables real-time assessment of whether the current generated sample incorporates the target factor, allowing for the generation of musically aligned samples without a costly retraining or fine-tuning process. We also propose to weigh all heads as a function of their learned probe performance, removing the need for empirical head selection tuning. The overall idea of the self-monitoring process is outlined in Fig.~\ref{fig:overall}.

The remainder of the paper is organized as follows. In Section~\ref{sec:understand_musicgen}, we evaluate whether the internal representations from individual attention heads of the pre-trained MusicGen model are suitable for downstream music classification tasks by training simple linear classifier probes. Then, in Section~\ref{sec:smitin}, we describe SMITIN, where these classifier probes are used to build custom ``knobs'' for inference-time control of MusicGen. In Section~\ref{sec:experiments}, we assess our proposed approach both objectively and subjectively in terms of intervention success and overall audio quality. Furthermore, we study in Section~\ref{sec:analysis} several important practical aspects of our approach, such as the number of audio examples required to obtain classifier probes sufficient for intervention, whether multiple interventions can be used simultaneously, and the ability of the intervention to prevent the autoregressive generative process from diverging over time into unrealistic and temporally incoherent music. In the Appendix, we additionally provide probing experiments on more downstream tasks and additional details of our subjective and objective results. Our source code will be available online\footnote{\url{https://github.com/merlresearch/smitin}}.

\begin{figure}[t]
     \centering
     \includegraphics[width=0.48\textwidth]{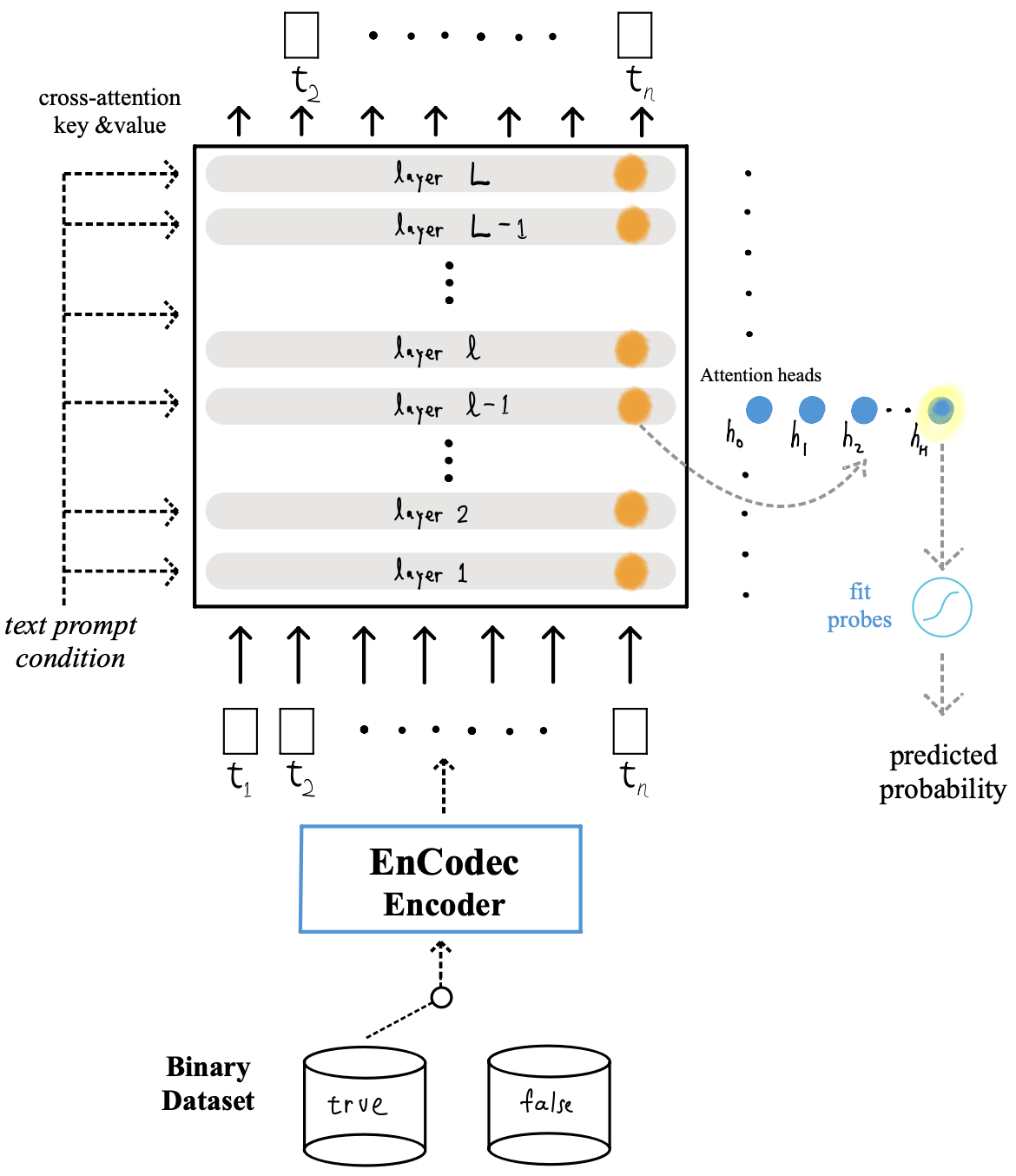}
     \caption{Overview of probing MusicGen. Each audio sample in a labeled dataset is converted to EnCodec tokens and input into MusicGen to predict the next token. 
     The activations for the last time step (orange dots) for each attention head in each layer (blue dots) are used to train a logistic regression classifier (probe).
     }
    \label{fig:probing_overview}
    \vspace{-3mm}
\end{figure}

\vspace{-.2cm}
\section{Understanding MusicGen}
\label{sec:understand_musicgen}

In this section, we seek to investigate and quantify the comprehension of music by each attention head within MusicGen~\cite{musicgen}. MusicGen is a pre-trained and publicly accessible generative music transformer that uses an EnCodec encoder~\cite{defossez2022high} to create discrete audio tokens, an autoregressive transformer to predict the next token, and an EnCodec decoder to output an audio signal. This analysis will provide insights into the model's potential for fine-grained control via attention-head steering. 
We use a probing task designed to assess the model's ability to distinguish musical pieces based on the presence or absence of specific instruments, and explore further downstream tasks in the Appendix.

\subsection{Autoregressive Transformer Models}

Architectures such as  MusicGen are characterized by the autoregressive generation of a sequence of audio frames by transformer models. They include a collection of $L$ multi-head self-attention layers (residual connections, normalization layers, and fully-connected feed-forward layers are employed as usual and not described here). At current time step $t$ in the generated sequence, the $l$-th self-attention layer %
computes $H$ self-attention heads $z_{l,h}(t)\in \mathbb{R}^D$ from an input vector $x_l(t) \in \mathbb{R}^{DH}$ as 
\begin{equation}
z_{l,h}(t) = \Att(W^{Q}_{l,h}x_l(t),W^{K}_{l,h}x_l(1\!:\!t),W^{V}_{l,h}x_l(1\!:\!t)), \label{eq:attentionvector}
\end{equation}
where $x_l(1\!:\!t)=[x_l(1),\dots,x_l(t)]$, $W^{Q}_{l,h}$, $W^{K}_{l,h}$, and  $W^{Q}_{l,h}$ denote the head-specific query, key, and value projection matrices, all in $\mathbb{R}^{D \times DH}$, and $\Att$ denotes the attention operator \cite{vaswani2017attention}.
The output vector $y_l(t) \in \mathbb{R}^{DH}$ is obtained by projecting back each head into $DH$-dimensional space and summing:
\begin{equation}
y_l(t) = \sum\limits^H_{h=1} W^O_{l,h} z_{l,h}(t), \label{eq:mhalayer}
\end{equation}
where $W^O_{l,h}\in\mathbb{R}^{DH \times D}$ is a projection matrix.

\subsection{Probing MusicGen}
\label{subsec:probing_musicgen}

We describe the methodology of probing MusicGen by evaluating the capability of its self-attention heads in recognizing instruments (i.e., determining whether a target instrument is present in the audio stream). The overview of the probing procedure is shown as Fig.~\ref{fig:probing_overview}. We create a dataset by curating data from MUSDB \cite{musdb18-hq} and MoisesDB \cite{moisesdb}, which offer multi-track recordings with isolated instrument stems. 
For a given target stem, we form a positive class of mixtures where the target stem is present, and a negative class of corresponding mixtures with the target stem removed as follows. %
First, we remove from every multi-track recording the time regions where the target stem is silent. Then, out of this pruned recording, the mixture of all of its stems is added to the positive class data, while the mixture of all of its stems except its target stem is added to the negative class data.
Subsequently, we process 3-second-long segments of these tracks for training (and testing), passing them through MusicGen to extract the intermediate activation $z_{l,h}(T)$ at the last time step $T$ for every self-attention layer $l$ and head $h$. 
This forms the basis for the training (and testing) sets of the probe classifier, wherein a simple logistic regression model is employed to distinguish the presence of the instrument.

\begin{figure}[t]
     \centering
     \subfigure[\textit{drums} (94.3\% / 0.903)]{\includegraphics[width=0.23\textwidth]{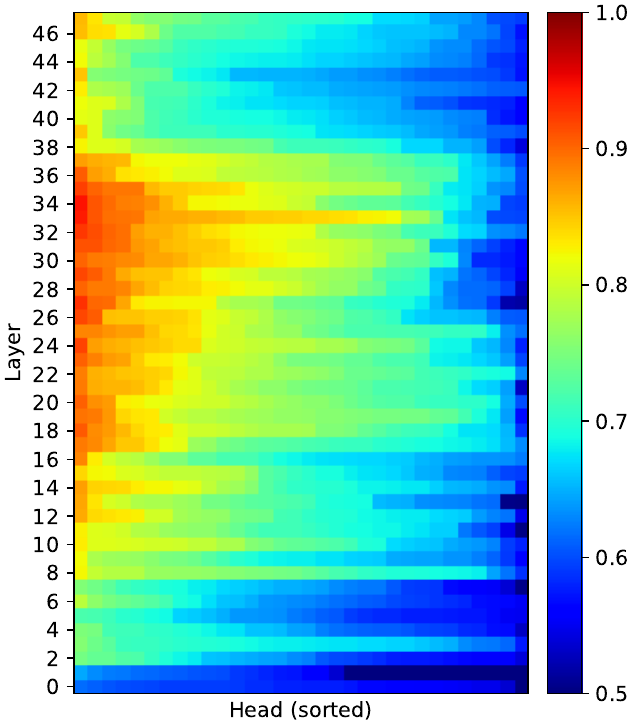}
     \label{subfig:probe_acc_drums}}
     \hfill
     \subfigure[\textit{bass} (89.1\% / 0.863)]{\includegraphics[width=0.23\textwidth]{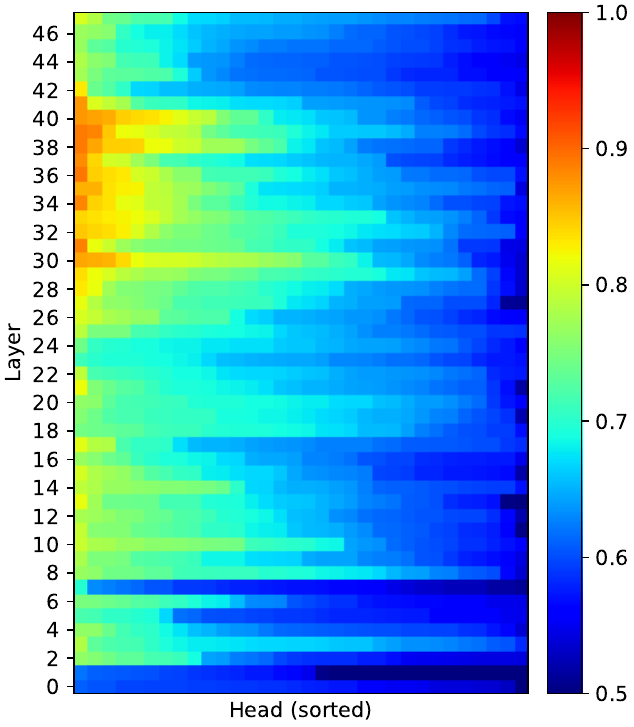}
     \label{subfig:probe_acc_bass}}
     \subfigure[\textit{guitar} (81.8\% / 0.787)]{\includegraphics[width=0.23\textwidth]{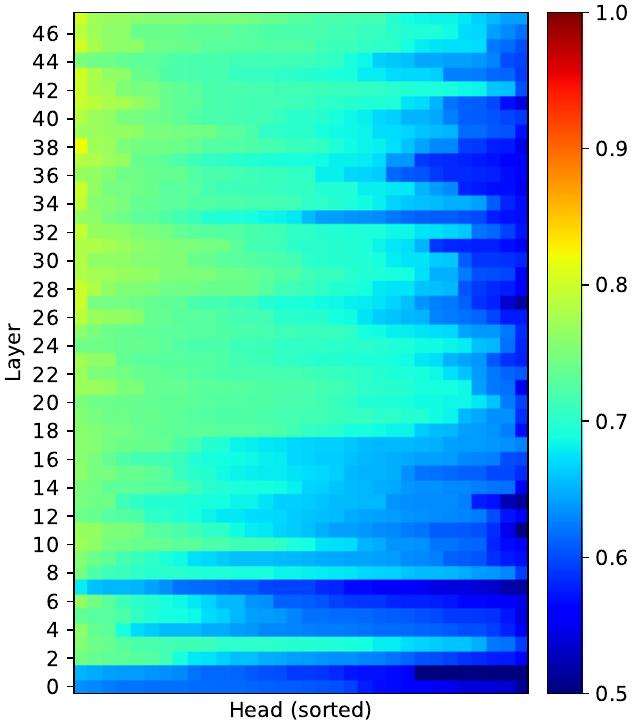}
     \label{subfig:probe_acc_guitar}}
     \hfill
     \subfigure[\textit{piano} (75.3\% / 0.712)]{\includegraphics[width=0.23\textwidth]{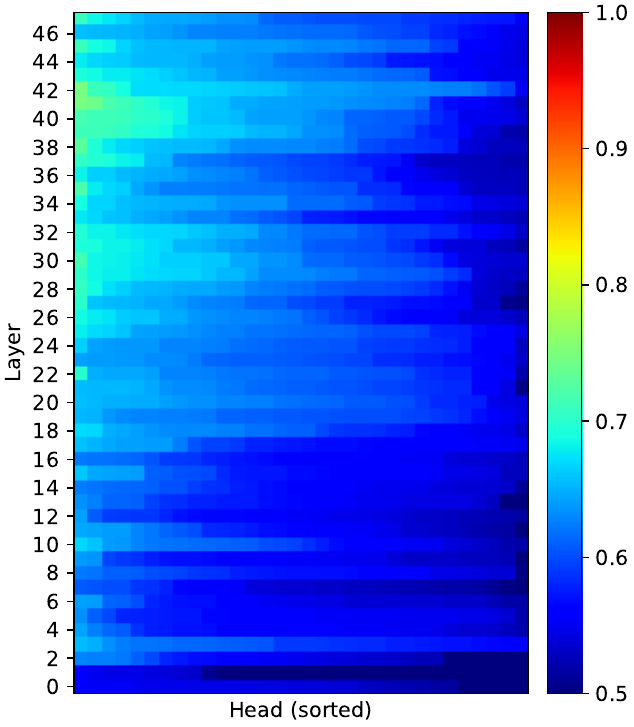}
     \label{subfig:probe_acc_piano}}
     \caption{Instrument recognition performance of individual attention head probes from the MusicGen$_\text{large}$ model activations, sorted by accuracy, with all colorbars normalized to the same range. The values in brackets indicate the highest accuracy of the probe classifier for each respective instrument task, followed by the threshold value $\tau$, which is defined in Section~\ref{subsec:self_monitoring}.
     }
    \label{fig:probe_acc_indiv_inst}
    \vspace{-3mm}
\end{figure}

The testing accuracy of probes from MusicGen$_\text{large}$ (3.3B parameters) across all self-attention layers $l$ and heads $h$ is illustrated in Fig.~\ref{fig:probe_acc_indiv_inst}. We observe that specific subsets of heads outperform others in detecting the presence of each target stem. While certain attention layers show better performance, it is notable that not all heads within each layer result in uniform performance; rather, their effectiveness varies considerably. This variation underscores the utility of head-wise probing in achieving precise control over the transformer's behavior. Furthermore, MusicGen's proficiency varies across instruments; it demonstrates a strong understanding of drums and bass, whereas its accuracy on guitar and piano is comparatively lower. This discrepancy suggests a potential bias in MusicGen's training dataset (see~\cite{musicgen}, Fig.~A.3).

\section{Self-Monitored Inference-Time Intervention}\label{sec:smitin}

\subsection{Inference-time Intervention (ITI)}
\label{subsec:ititheory}

Reference \cite{iti} first recognized that the output of the model could be somewhat controlled at inference by intervening in (i.e., modifying) the computation of the output of the multi-head self-attention layer in Eq.~\eqref{eq:mhalayer}. This intervention is done by adding a term to the heads $z_{l,h}(t) \in \mathbb{R}^D$ before the projection:
\begin{equation}
y_l(t) = %
\sum\limits^H_{h=1} W^O_{l,h} \left(z_{l,h}(t) + \alpha w_{l,h} \cdot \sigma_{l,h} \theta_{l,h}\right),
\label{eq:mhalayerwiti}
\end{equation}
where $\theta_{l,h} \in \mathbb{R}^D$ is a vector representing the head-specific ITI ``direction'' in the $D$-dimensional head-specific space, $\sigma_{l,h} \in \mathbb{R}^+$ is a head-specific normalization of $\theta_{l,h}$, $\alpha \in \mathbb{R}^+$ is the system-wide ITI strength, and $w_{l,h}\geq 0$ is a head-specific strength weighting.
While the formulations are equivalent, we depart somewhat from \cite{iti} by introducing $w_{l,h}$ as a separate term, whereas it was factored into $\theta_{l,h}$ in their notations. This helps to disambiguate the dual function played by $\theta_{l,h}$ in their formulation. 

The head-specific ITI directions $\theta_{l,h}$ are obtained through classifier probes. Following \cite{iti}, we find an auxiliary probing dataset suited to train a classifier mirroring our ITI goal (e.g., drum classifier if the goal is to add drums). We then run the dataset through our (frozen) generation network and collect a set of heads $z_{l,h}(t)$ for all $l,h$. For each $l$ and $h$, we then train a distinct logistic regression classifier probe with parameters $\tilde{\theta}_{l,h}$, whose prediction is obtained as $\sigmoid(\langle \tilde{\theta}_{l,h}, z_{l,h}(t) \rangle)$. Once probe training is complete, we set ITI direction $\theta_{l,h}$ as the final $\tilde\theta_{l,h}$, and $\sigma_{l,h}$ as the standard deviation of $\langle\theta_{l,h},z_{l,h}(t)\rangle$ for all $z_{l,h}(t)$ obtained on the combined probing training and testing data. We also take note of the final classifier accuracy $\acc_{l,h}$ on the probing test data. In the case of MusicGen, we apply the generative model in unconditional generation mode for probing, as our probing dataset generally lacks text queries for its audio samples.

In \cite{iti}, head-specific weights $w_{l,h}$ are set by finding the top-$K$ heads in terms of classifier probe accuracy $\acc_{l,h}$, setting their $w_{l,h}$ to 1 and the others to 0. The most effective $\alpha$ and $K$ are found by hyperparameter grid search.

\subsection{Sparse intervention}
\label{subsec:sparse}

In music generation, the approach above may present some limitations. For example, for the task of audio continuation, we observe that ITI often leads to changes that are too abrupt to be musically plausible (see Section~\ref{subsec:eval_qual}). As mitigation, we propose to diminish the ITI frequency across time steps, potentially allowing the generation process to better align with the underlying rhythmic structure of the generated music. This corresponds to replacing the ITI weights $w_{l,h}$ in Eq.~\eqref{eq:mhalayerwiti} by time-varying weights $w_{l,h}(t)$, which can only be non-zero for $t=t_0+is,$ $i\in\mathbb{N}$, where $t_0$ is an intervention start time and 
$s$ represents the number of steps between each ITI (e.g., $s=5$ to perform ITI every 5 time steps). The value of $w_{l,h}(t)$ for $t=t_0+is$ can be set by another criterion, such as the original one based on top-$K$ heads.

\subsection{Soft-weighting}
\label{subsec:softweighting}

One crucial limitation of \cite{iti} is that $K$, the number of heads on which ITI is performed (i.e., the number of pairs $(l,h)$ for which $w_{l,h}(t) \neq 0$), is a hyperparameter that must be tuned. We additionally propose a hyperparameter-free soft-weighting approach based on the collected probe accuracies $\acc_{l,h}$, and show that it is sufficient to perform effective ITI (see Appendix Table~\ref{tab:ablation_topk}). 
In practice, we propose setting the weights as
\begin{equation}
w_{l,h}(t) = \left(\frac{\acc_{l,h} - \acc_\text{min}}{\acc_\text{max} - \acc_\text{min}}\right)^{\!c},
\label{eq:softweight}
\end{equation}
with $\acc_\text{min}$ and $\acc_\text{max}$ the minimum and maximum accuracies recorded across all $l$ and $h$, and $c$ a power factor allowing to modulate the relative weights of heads with accurate vs.\ inaccurate classifier probes (we use $c=3$). By construction, $w_{l,h}(t)$ is guaranteed to fall between 0 and 1. %

\subsection{Automated Intervention Modulation by Self-monitoring}
\label{subsec:self_monitoring}

Ideally, we would expect a system capable of modulating the ITI strength to be most effective, as systems with time-invariant $w_{l,h}(t)$ make no use of (and, as such, cannot react to) the state of the inference model during generation.

As our core contribution, we propose to use the classifier probes to drive such a modulation. 
We first define as $\cH_K$ the set of top-$K$ heads $(l,h)$ by probe classifier accuracy. For each new generation time step $t$, we collect the set $\cC(t)$ of output predictions of the classifier probes for all heads in $\cH_K$ before intervening on them, i.e.,
\begin{equation}
\cC(t)=\left\lbrace \sigmoid\left(\langle \theta_{l,h}, z_{l,h}(t)\rangle\right) \,|\, (l,h)\in \cH_K \right\rbrace.
\end{equation}
The statistics of $\cC(t)$ reflect the confidence of the probes regarding the success of the intervention. By comparing them to the baseline accuracies obtained on the probing training data, we can devise a modulation scheme to update $w_{l,h}(t)$. We compute the median $\bar{\cC}(t)=\med(\cC(t))$ and the change in this median since the prior generation step with intervention, $\Delta(t)= \bar{\cC}(t) - \bar{\cC}(t\!-\!s)$.
We then define a threshold based on the median and standard deviation of the set $\cA$ of probe accuracies $\acc_{l,h}$ on the probing training data over the same heads $\cH_K$, 
\begin{equation}
    \cA = \{\acc_{l,h} \,|\, (l,h)\in \cH_K\},
\end{equation}
setting the threshold as $\tau = \med(\cA)-\std(\cA)$.

The update algorithm is then defined following:
\begin{itemize}[leftmargin=*,noitemsep,topsep=0pt]
    \item For the first generation time step with ITI $t_0$, $w_{l,h}(t_0)$ is set following Eq.~\eqref{eq:softweight}, and we set $\Delta(t_0) \leftarrow 0$,
    \item For each subsequent generation time step with ITI $t\!+\!s$, we have 3 cases:
    \begin{enumerate}[leftmargin=*,noitemsep,topsep=0pt]
        \item if $\bar{\cC}(t) \!<\! \tau$ and $w_{l,h}(t) \!>\! 0$, we set 
            \begin{equation}
                w_{l,h}(t+s) \leftarrow w_{l,h}(t) \cdot (1 - \Delta(t)),
            \end{equation}
        \item if $\bar{\cC}(t) \!<\! \tau$ and $w_{l,h}(t) = 0$, we reset to the initial value
            \begin{equation}
                w_{l,h}(t+s) \leftarrow w_{l,h}(t_0),
            \end{equation}
        \item if $\bar{\cC}(t) \geq \tau$, we set $w_{l,h}(t+s) \leftarrow 0$.
    \end{enumerate}
\end{itemize}

\section{Controlled Experiments on Instrument Addition}
\label{sec:experiments}

Here, we validate the effectiveness of our proposed contribution by presenting a thorough investigation of ITI applied to the music generative task of instrument addition, a task for which evaluation is relatively straightforward.
This section outlines methods to quantify the effectiveness of ITI techniques objectively, complemented by a subsequent subjective evaluation.

\subsection{Setup}
\label{subsec:setup}

For the evaluation of ITI on MusicGen, we utilize the \textit{large mono} configuration, which is composed of 48 layers with 32 attention heads each, amounting to a total of 1,536 heads. Unless otherwise noted, the self-monitoring configuration of SMITIN is set to an intervention strength of $\alpha=5.0$, a sparse intervention rate of $s=5$, and $K=16$ for the selection of monitoring probes.

For the instrument addition task, we curate probing datasets for drums, bass, guitar, and piano detection from MUSDB \cite{musdb18-hq} and MoisesDB \cite{moisesdb}, both of which provide multi-track recordings with isolated instrument stems. Stems from their training (resp.\ testing) partitions are used to form the probing training (resp.\ testing) partition. %
To create positive and negative classes, i.e., one including and one excluding the target instrument, we proceed as follows. We collect a first set consisting of the mixed audio from each track recording after removing the target instrument stem, and trim silent sections. We then collect a second set consisting of the target instrument stems, and trim silent sections. We then generate negative samples using audio segments randomly sampled from the first set, and positive samples by randomly mixing a mixture without the target instrument (from the first set) with a stem of the target instrument (from the second set). Following silence trimming, the datasets amount to 8.3, 8.0, 13.1, and 5.7 hours for drums, bass, guitar, and piano, respectively. These durations represent the total amount of paired data available for probing. Since the negative samples are more abundant, we balanced the dataset to ensure an equal number of positive and negative samples, based on the total duration of positive samples.

\vspace{-.7cm}
\subsubsection{Generation Approach}
\label{subsubsec:generation_approach}

Our experimental setup includes two contexts in which to perform ITI: audio continuation and text-to-music. In both cases, the objective is to perform the task while adding the target instruments (drums, bass, guitar, and piano) into the music pieces through ITI. 

\noindent\textit{Audio continuation}: Given an input music signal, we continue to generate a music piece while trying to add the target instrument. The input segment is a 3-second-long music segment obtained from the MUSDB and MoisesDB test datasets, which does not contain the target instrument.

\noindent\textit{Text-to-music}:  This task uses text prompts from the MusicCaps \cite{musiccaps_musiclm} dataset to guide the generation process, allowing us to evaluate the effectiveness of ITI in the context of text-conditioned generation. We use the \text{aspect list} (e.g., \textit{``pop, tinny wide hi hats, mellow piano melody"}) over the \text{free text caption} to further investigate the impact of adding simple instructions, e.g., ``add inst." to the tag list.

\vspace{-.5cm}

\subsubsection{Objective Metrics}
\label{subsubsec:obj_metric}

Our objective evaluation comprises three metrics designed to assess the effectiveness of the ITI process. 
We generate 1,000 music pieces, each 30 seconds long, for each experimental configuration.

\textbf{Success Rate}: 
We introduce Success Rate to quantify the efficacy of ITI in successfully adding the desired instrument to the music. This metric is derived from our self-monitoring technique from Section~\ref{subsec:self_monitoring}, which measures the likelihood of the target musical factor based on the timewise probability inferred by the top-$K$ best-performing probes, where $K=16$. For each audio, the success rate is calculated as $n_\text{success} / N$, where $n_\text{success}$ is the number of samples $t$ after the start of the intervention for which $\bar{\cC}(t) > \tau$, and $N$ is the total number of time steps generated after the intervention. The reliability of this metric is contingent upon the top-$K$ probes' test performance in the probing task.

Although the probing technique performs well on real-world audio samples, as demonstrated in Section~\ref{sec:understand_musicgen}, there is some uncertainty regarding its effectiveness on generated samples. Also, because the Success Rate metric uses the same internal monitoring probes of MusicGen as the proposed intervention techniques, there is some concern that it may overestimate success in a way other instrument recognition models would not. Thus, to further validate the Success Rate, we compared its results on all the generated samples used for subjective evaluation in Section~\ref{subsec:eval_sbj} against decisions from other instrument recognition models. Specifically, we measured the Spearman correlation between the logits produced by our internal probes and those produced by external models such as YAMNet~\cite{yamnet}, PaSST~\cite{passt}, CLAP~\cite{l-clap}, and ConvNeXt~\cite{liu2022convnet}. YAMNet and PaSST were both trained on AudioSet~\cite{audioset}, with PaSST further fine-tuned using the OpenMIC dataset~\cite{openmic}. For CLAP-based recognition, we trained a two-layer MLP to classify the embeddings generated by the frozen CLAP encoder, training on the same dataset as our internal probes. ConvNeXt was trained and tested using the same dataset as well (see Appendix A1 for detailed setup). %
We used publicly-available weights for YAMNet\footnote{\scriptsize{\url{https://github.com/tensorflow/models/tree/master/research/audioset/yamnet}}}, 
PaSST\footnote{\scriptsize{\url{https://github.com/kkoutini/PaSST}}}, and CLAP\footnote{\scriptsize{\url{https://github.com/LAION-AI/CLAP}}}.

The results in Table~\ref{tab:obj_correlation} demonstrate a strong positive correlation, confirming that our internal probes align with external models in recognizing the desired musical traits. Furthermore, as we show in the analysis of subjective evaluation in Section~\ref{subsec:eval_sbj}, all the instrument recognition methods also correlate well with human decisions, providing additional validation of the models’ alignment with perceptual recognition.

A key advantage of using the Success Rate metric with MusicGen’s internal probes is that it allows for frame-wise recognition during generation, offering precise detection of musical traits throughout the entire sample. For example, consider a 10-second music sample where drums are only present in the last 3 seconds. External instrument recognition models, which often operate with larger window sizes, may struggle to accurately identify that drums are only present in 30\% of the sample. In contrast, our internal probes can precisely distinguish such temporal details, identifying the proportion of the sample that contains the target instrument. This level of granularity is particularly important in controlled generation tasks, where it is critical to accurately capture the presence and distribution of musical traits in the generated audio.

\begin{table}[t]
\centering
\sisetup{
    detect-weight, %
    mode=text, %
    tight-spacing=true,
    round-mode=places,
    round-precision=3,
    table-format=1.3,
    table-number-alignment=center
    }
    \vspace{-.2cm}
\caption{
    Spearman correlation between the Success Rate and other instrument recognition methods. All p-values are below 1e-5.
}
\label{tab:obj_correlation}
\setlength\tabcolsep{5.0pt}
\footnotesize
\begin{tabular}{l*{5}{c}}
\toprule
 & \multicolumn{5}{c}{\textbf{Correlation Coefficient $\rho$ ($\uparrow$)}} \\ 
\cmidrule(lr){2-6}
 \textbf{Method}                    & \textit{drums} & \textit{bass} & \textit{guitar} & \textit{piano} & \textit{avg.} \\ 
\midrule
\textbf{YAMNet}~\cite{yamnet}      & 0.71            & 0.50           & 0.58             & 0.33            & 0.53           \\
\textbf{PaSST}~\cite{passt}       & 0.69            & 0.73           & 0.63             & 0.40            & 0.63           \\
\textbf{CLAP}~\cite{l-clap}        & 0.82            & 0.74           & 0.68             & 0.66            & 0.75           \\
\textbf{ConvNeXt}~\cite{liu2022convnet}    & 0.73            & 0.84           & 0.70             & 0.74            & 0.75           \\
\bottomrule
\end{tabular}
\end{table}

\textbf{Fréchet Audio Distance (FAD)}: FAD \cite{fad} is used to compare generated music with real music datasets, which helps ensure that the intervention does not cause an unrealistic shift in MusicGen's output distribution. Instead of the conventional approach of using VGGish \cite{vggish} as the deep encoder, we use L-CLAP mus \cite{l-clap} for dimension reduction, as it has been shown to offer a more accurate representation of music \cite{fad_analysis}. As the real music datasets, we adopt MUSDB (drums/bass) or MoisesDB (guitar/piano) for audio continuation and MusicCaps for text-to-music generation.

\textbf{Similarity Measurement in audio continuation}: For the audio continuation task, we measure the musical similarity between input and generated samples. We employ the Music Effects Encoder \cite{mee}, a model trained with self-supervised contrastive learning designed to encode song identity. The encoder is used to embed both input and generated samples, and their cosine similarity is calculated. This metric assesses how well the identity of the input music is preserved in the generated output.

\subsection{Qualitative Evaluation}
\label{subsec:eval_qual}
\begin{figure*}[th]
    \centering
    \begin{minipage}{0.44\textwidth}
        \centering
        \includegraphics[width=\textwidth]{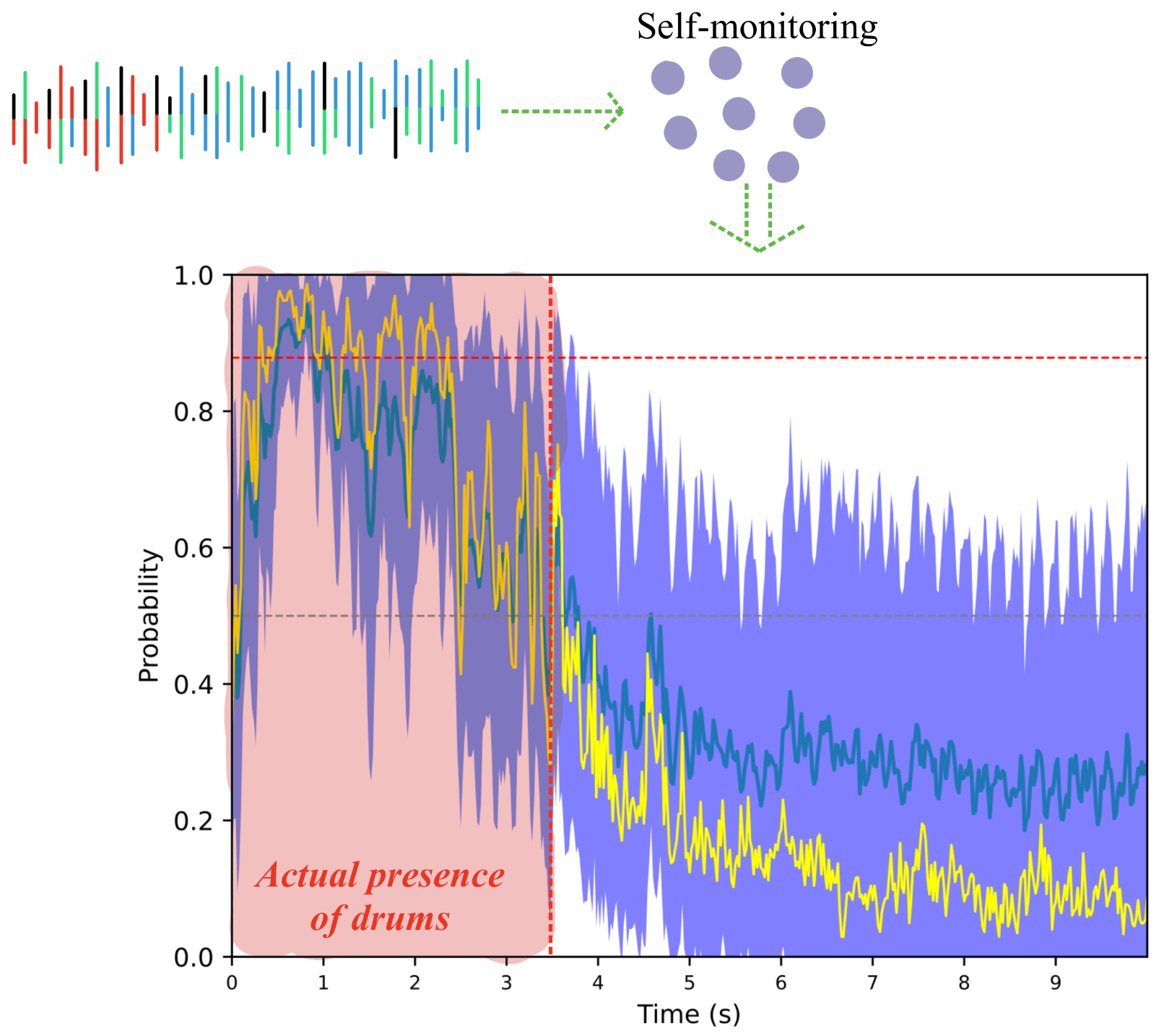}
        \label{fig:large_figure}
    \end{minipage}
    \hfill
    \begin{minipage}{0.55\textwidth}
        \centering
        \includegraphics[width=0.45\textwidth]{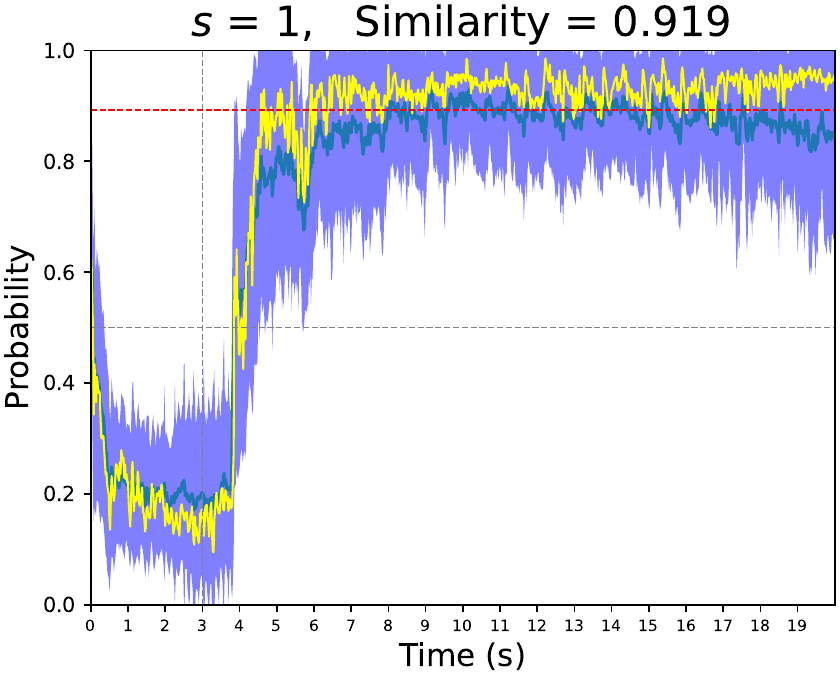}
        \includegraphics[width=0.45\textwidth]{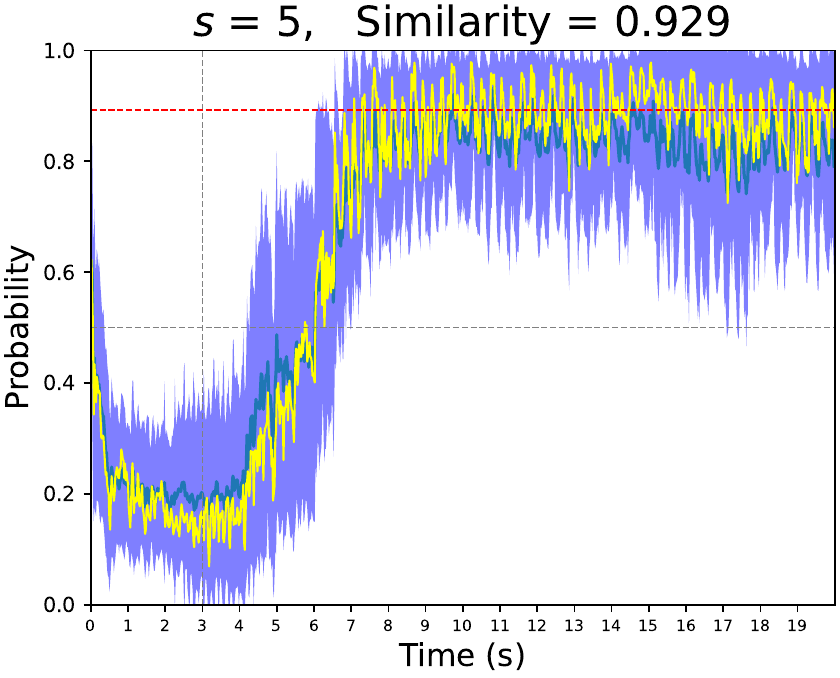}
        \includegraphics[width=0.45\textwidth]{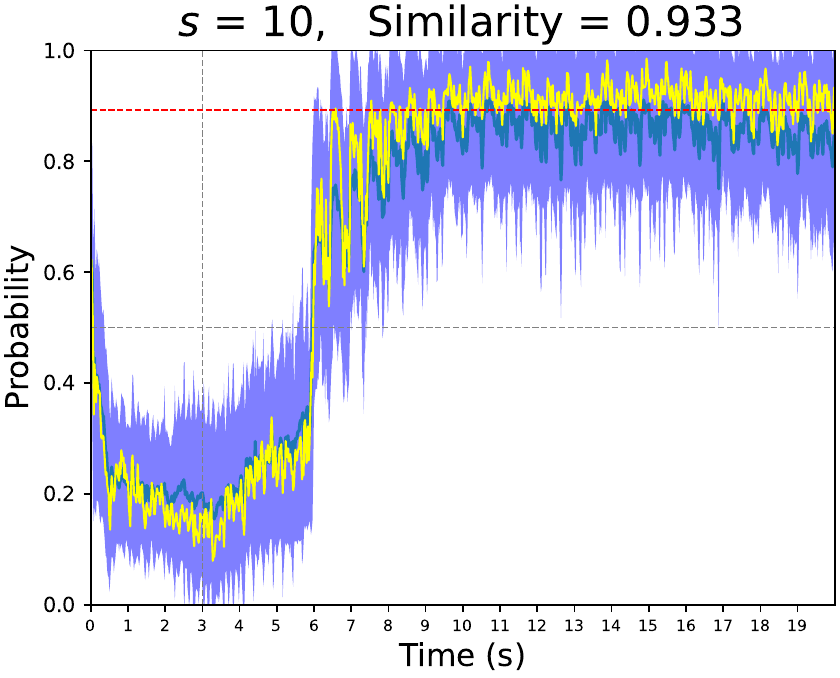}
        \includegraphics[width=0.45\textwidth]{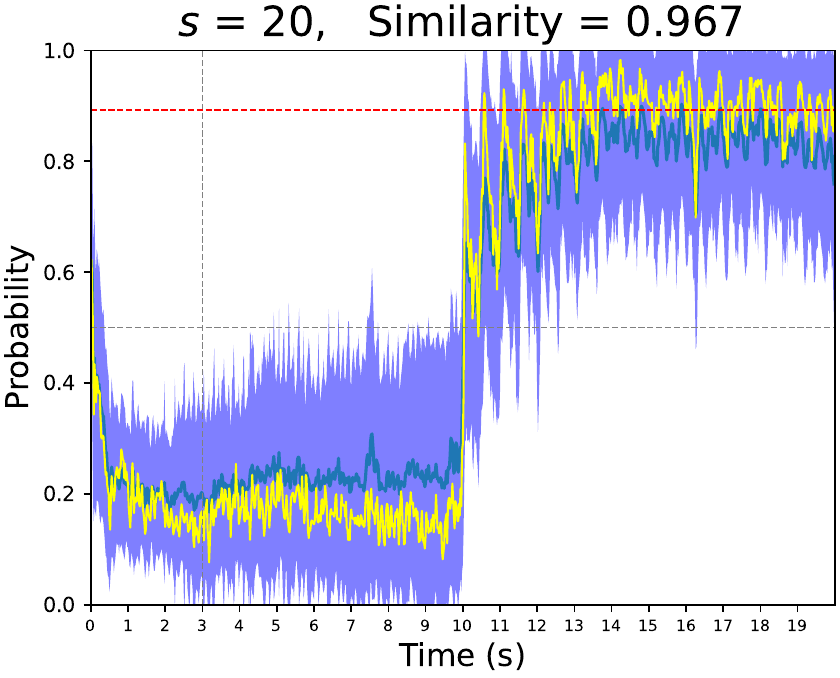}
    \end{minipage}
    \caption{Inferred prediction of the top-K probes' monitored decision along the time axis. The yellow line, green line, and the shaded blue region denote the median, mean, and standard deviation of inferred outputs by the probes, respectively. The red dashed line indicates threshold value $\tau$ of the current monitoring probes. (Left) Monitored result on a real-world music sample. The high prediction (close to 1.0) until 3.5 seconds reflects the actual presence of drums, which aligns with the audio sample where drums are present only up to that point. (Right) The four sub-plots display the results of audio continuation on the same input music with varying ITI frequencies ($s=[1, 5, 10, 20]$). These illustrate that more frequent intervention leads to a swifter convergence towards the target musical factor, at the expense of losing musical consistency with the input music.}
    \label{fig:qual_probe_monitoring}
    \vspace{-.3cm}
\end{figure*}

We qualitatively analyze the performance of top-$K$ probes' monitoring capability through Fig.~\ref{fig:qual_probe_monitoring}. The red dashed line in each plot indicates the threshold value $\tau$, as mentioned in Section~\ref{subsec:self_monitoring}. From the left side of the figure, it is apparent that the median prediction (yellow line) more closely aligns with the actual presence of the target instrument than the mean prediction (green line). Consequently, we use $\bar{\cC}(t)=\med(\cC(t))$ to monitor during inference time.

The sub-plots on the right illustrate how the hyper-parameter $s$ influences the outcome of the audio continuation task. We observe a relationship between the intervention frequency $s$ and the rapidity with which the model aligns toward the target factor. 
This suggests that more frequent interventions cause the transformer to prioritize the target factor more heavily in subsequent token generation, potentially at the expense of referencing the similarity to previous musical context. We discuss this trade-off between intervention quantity and musical coherence throughout this section.

\subsection{Objective Evaluation}
\label{subsec:eval_obj}
Utilizing the metrics introduced in Section~\ref{subsubsec:obj_metric}), we objectively evaluate our methods on \textit{audio continuation} and \textit{text-to-music}. We compare with baseline methods that do not incorporate ITI, such as MusicGen's standard unconditional and text-prompt conditioned generations. Additionally, we benchmark against the original ITI method, which applies a uniform intervention at every time step (i.e., $s=1$) with a configuration of $\alpha=5.0$ and $K=64$ to apply the intervention with the equal amount $\alpha$ to each of the $K$ selected heads. For \textit{audio continuation}, we examine another ITI approach, weight decay, where $\alpha$ is progressively reduced across time steps during the generation process. For \textit{text-to-music}, our evaluation includes both our default configuration ($\alpha=5.0$) and a stronger intervention level of $\alpha=10.0$ to observe the trade-off between the intervention strength and the preservation of musical fluency within the generated samples.

\begin{table*}[t]
\centering
\sisetup{
    detect-weight, %
    mode=text, %
    tight-spacing=true,
    round-mode=places,
    round-precision=3,
    table-format=1.3,
    table-number-alignment=center
    }
    \vspace{-.2cm}
\caption{
    Objective evaluation on \textit{audio continuation}. FAD$_\text{L-CLAP mus}$ is computed with MUSDB (drums/bass) and MoisesDB (guitar/piano).
}
\setlength\tabcolsep{3.5pt}
\footnotesize
\resizebox{1.\linewidth}{!}
{
\begin{tabular}{cl*{5}{S[round-precision=1,
    table-format=2.1]}*{5}{S}*{5}{S}}
\toprule
\multicolumn{2}{c}{\textbf{Method}}
& \multicolumn{5}{c}{\textbf{Success Rate [\%] ($\uparrow$)}}                                       & \multicolumn{5}{c}{\textbf{FAD$_\text{L-CLAP mus}$ ($\downarrow$)}}  &  \multicolumn{5}{c}{\textbf{Similarity ($\uparrow$)}}   \\ 
\cmidrule(lr){1-2}\cmidrule(lr){3-7}\cmidrule(lr){8-12}\cmidrule(lr){13-17}
\textbf{ITI}                & \textbf{Configuration}                   & \textit{drums} & \textit{bass} & \textit{guitar} & \textit{piano} & \textit{avg.} & \textit{drums} & \textit{bass} & \textit{guitar} & \textit{piano} & \textit{avg.} & \textit{drums} & \textit{bass} & \textit{guitar} & \textit{piano} & \textit{avg.}  \\ 
\midrule
\multirow{2}{*}{\textit{\sffamily X}}  & \textbf{unconditioned}                                       & 12.5          & 13.5         & 3.6           & 0.0          & 7.4         & 0.326          & 0.253         & 0.377           & 0.320          & 0.319            & 0.864          & 0.933         & 0.929           & 0.941          & 0.916         \\
& \textit{\textbf{``add \textless{}inst.\textgreater{}''}} & 16.6          & 27.4         & 5.5           & 1.7          & 12.8         & 0.364          & 0.301         & 0.385           & 0.351          & 0.350        & 0.833          & 0.905         & 0.929           & 0.936          & 0.900         \\ 
\cdashlinelr{1-17}
\multirow{4}{*}{\textbf{\checkmark}} & \textbf{original ITI}                                      & 21.3          & 76.0         & 17.2           & 85.2          & 49.9         & 0.358          & 0.510         & 0.447           & 0.368          & 0.420       & 0.868          & 0.858         & 0.919           & 0.939          & 0.896         \\
& \textbf{weight decay ITI}                                  & 13.2          & 17.5         & 15.4           & 3.1          & 12.3         & 0.336          & 0.259         & 0.378           & 0.337          & 0.327         & 0.871          & 0.927         & 0.933           & 0.941          & 0.918         \\
& \textbf{SMITIN}                                      & 20.6          & 30.4         & 33.8           & 8.7          & 23.3         & 0.346          & 0.267         & 0.397           & 0.337          & 0.336          & 0.857          & 0.922         & 0.935           & 0.938          & 0.913         \\
& \textbf{SMITIN + \textit{``add \textless{}inst.\textgreater{}"}}                               & 16.9          & 28.0         & 16.9           & 17.4          & 19.8         & 0.381          & 0.303         & 0.391           & 0.374          & 0.362        & 0.839          & 0.909         & 0.929           & 0.942          & 0.904         \\ 
\bottomrule
\end{tabular}
}
\vspace{-2mm}
\label{tab:obj_ITI_audio_cont}
\end{table*}

Table~\ref{tab:obj_ITI_audio_cont} presents the objective evaluation for the \textit{audio continuation} task. The baseline methods reveal that MusicGen inherently tends to add certain instruments without explicit direction with varying success rates across instruments; drums and bass are more easily added compared to guitar and piano. Intriguingly, text-prompted generation does not consistently achieve the target instruction.
Intervention methods, on the other hand, consistently outperform baseline approaches in successfully directing MusicGen toward adding the desired instruments, as indicated by the success rate. The original ITI, applying a constant level of intervention at each generation step, leads to a higher success rate but significantly alters the distribution of generated content, which can be observed via the similarity score for maintaining musical consistency. In contrast, the weight decay ITI approach has a subtler influence on the quality but produces a lower success rate. Our proposed SMITIN shows a notable 10.5\% jump over text-prompt conditioning and is better at retaining the model's output distribution and generating consistent music. When SMITIN is deployed in conjunction with text conditioning, it outperforms text-based instructions, but not to the same extent as when used independently (likely because we do not intervene on the text cross-attention layers). %

\begin{table*}[t]
\centering
\sisetup{
    detect-weight, %
    mode=text, %
    tight-spacing=true,
    round-mode=places,
    round-precision=3,
    table-format=1.3,
    table-number-alignment=center
    }
\caption{
    Objective evaluation on \textit{text-to-music}. FAD$_\text{L-CLAP mus}$ is computed with MusicCaps.
}
\setlength\tabcolsep{3.5pt}
\footnotesize
\begin{tabular}{cc*{5}{S[round-precision=1,
    table-format=2.1]}*{5}{S}}
\toprule
\multicolumn{2}{c}{\textbf{Method}}
& \multicolumn{5}{c}{\textbf{Success Rate [\%] ($\uparrow$)}}                                       & \multicolumn{5}{c}{\textbf{FAD$_\text{L-CLAP mus}$ ($\downarrow$)}}   \\ 
\cmidrule(lr){1-2}\cmidrule(lr){3-7}\cmidrule(lr){8-12}
\textbf{ITI}                & \multicolumn{1}{l}{\textbf{Configuration}}                & \textit{drums} & \textit{bass} & \textit{guitar} & \textit{piano} & \textit{avg.} & \textit{drums} & \textit{bass} & \textit{guitar} & \textit{piano} & \textit{avg.} \\ 
\midrule
\multirow{2}{*}{\textit{\sffamily X}} & \multicolumn{1}{l}{\textbf{text}}                                         & 29.9                              & 50.4                             & 17.4                               & 7.1                              & 26.2                             & \multicolumn{5}{c}{0.482 (indep.\ of inst.)}     \\
& \multicolumn{1}{l}{\textbf{text + \textit{``add \textless{}inst.\textgreater{}"}}} & 34.4                              & 56.1                             & 22.5                               & 10.5                              & 30.8                             & 0.474                              & 0.481                             & 0.473                               & 0.488                              & 0.479     \\ 
\cdashlinelr{1-12}
\multirow{3}{*}{\textbf{\checkmark}} & \multicolumn{1}{l}{\textbf{text + original ITI}}                      & 44.3                              & 79.6                             & 33.4                               & 51.0                              & 52.0                             & 0.506                              & 0.577                             & 0.555                               & 0.517                              & 0.538                       \\
& \multicolumn{1}{l}{\textbf{text + SMITIN ($\alpha=5.0$)}}                      & 29.9                              & 53.4                             & 23.1                               & 12.5                              & 29.7                             & 0.471                              & 0.493                             & 0.499                               & 0.487       & 0.487                  \\
& \multicolumn{1}{l}{\textbf{text + SMITIN ($\alpha=10.0$)}}                     & 40.9                              & 60.7                             & 35.4                               & 21.0                              & 39.5                             & 0.481                              & 0.516                             & 0.507                               & 0.485                              & 0.497                                 \\ 
\bottomrule
\end{tabular}
\vspace{-3mm}
\label{tab:obj_ITI_txt_cond}
\end{table*}

For the \textit{text-to-music} task, we explore the interplay of ITI with text prompts, as shown in Table~\ref{tab:obj_ITI_txt_cond}. In this context, we utilize MusicCaps text prompts as the basis of MusicGen's performance. Consistent with earlier observations, the original ITI method, despite showing the highest success rate in instrument addition, performs the lowest in terms of FAD, indicating a more pronounced shift from the natural music distribution. SMITIN, in its base configuration ($\alpha=5.0$), achieves an average performance on par with additional text guidance. We further analyze SMITIN with an increased intervention strength of $\alpha=10.0$ and observe that a clear trade-off emerges between the success rate and the naturalness of the generated music. This outcome highlights the flexibility of SMITIN, offering users a tunable ``knob" to balance between precision in achieving specific musical characteristics and maintaining the authenticity of the musical piece.

\begin{table}[t]
\centering
\sisetup{
detect-weight, %
mode=text, %
tight-spacing=true,
round-mode=places,
round-precision=2,
table-format=3.2,
table-number-alignment=center
}
\vspace{-.24cm}
\caption{Average and 95\% confidence interval of subjective listening tests for overall audio quality across four instrument addition interventions for audio continuation and text-to-music.
}
\vspace{.05cm}
\label{tab:subjective_small}
\footnotesize
\begin{tabular}{l*{2}{S@{\,\( \pm \)\,}S[table-format=1.2]}}
\toprule
\textbf{Method} & \multicolumn{2}{c}{\textbf{Continuation}}      & \multicolumn{2}{c}{\textbf{Text-to-Music}}      \\ \midrule
\textbf{No intervention}      &  3.35 & 0.10   & 3.67 & 0.09 \\
\textbf{\textit{``add \textless{}inst.\textgreater{}"}}       & 3.27 & 0.10  & 3.61 & .09 \\ 
\textbf{original ITI}      &  3.28 & 0.11   & 3.55 & 0.09 \\
\textbf{SMITIN}      &  3.34 & 0.10   & 3.80 & 0.09 \\\midrule
\textit{MoisesDB}    &  \multicolumn{4}{c}{\textit{3.98 $\pm$ 0.29}} \\ \bottomrule
\end{tabular}
\vspace{-.4cm}
\end{table}
\begin{table}[t]
\centering
\sisetup{
detect-weight, %
mode=text, %
tight-spacing=true,
round-mode=places,
round-precision=2,
table-format=1.2,
table-number-alignment=center
}
\caption{Subjective results across multiple algorithms, comparing human (\textit{Hum.}) detection on the presence of instruments added by intervention, with Success Rate (\textit{Alg.}).  All Spearman correlation coefficients ($\rho$) have p-values less than 1e-5.
}
\label{tab:subjective_presence}
\footnotesize
\setlength{\tabcolsep}{3pt}
\resizebox{1.\linewidth}{!}
    {
\begin{tabular}{l*{5}{S[table-format=1.2]}*{5}{S}}
\toprule
 & \multicolumn{5}{c}{\textbf{Audio continuation}} & \multicolumn{5}{c}{\textbf{Text-to-music}}  \\
\cmidrule(lr){2-6}\cmidrule(lr){7-11}
 & & \multicolumn{2}{c}{\textbf{Top}} & \multicolumn{2}{c}{\textbf{Bottom}} &  & \multicolumn{2}{c}{\textbf{Top}}  & \multicolumn{2}{c}{\textbf{Bottom}} \\
\cmidrule(lr){3-4}\cmidrule(lr){5-6}\cmidrule(lr){8-9}\cmidrule(lr){10-11}
\textbf{Inst.}& $\rho$ & \textit{Hum.} & \textit{Alg.} & \textit{Hum.} & \textit{Alg.} & $\rho$ & \textit{Hum.} & \textit{Alg.} & \textit{Hum.} & \textit{Alg.} \\\midrule

drums   & 0.76 & 0.88 & 0.81 & 0.19 & 0.13 & 0.80 & 0.95 & 0.95 & 0.16 & 0.18 \\
bass    & 0.42 & 0.78 & 0.89 & 0.48 & 0.06 & 0.58 & 0.80 & 0.96 & 0.38 & 0.13 \\ 
guitar  & 0.40 & 0.78 & 0.76 & 0.52 & 0.15 & 0.67 & 0.86 & 0.88 & 0.34 & 0.14 \\ 
piano   & 0.32 & 0.60 & 0.72 & 0.33 & 0.32 & 0.55 & 0.70 & 0.79 & 0.31 & 0.23 \\ 
avg.    & 0.48 & 0.76 & 0.80 & 0.39 & 0.17 & 0.67 & 0.83 & 0.90 & 0.30 & 0.17 \\ \bottomrule
\end{tabular}
}
\vspace{-.3cm}
\end{table}

\begin{table}[t]
\centering
\sisetup{
    detect-weight, %
    mode=text, %
    tight-spacing=true,
    round-mode=places,
    round-precision=3,
    table-format=1.3,
    table-number-alignment=center
    }
\caption{
    Spearman correlation of instrument recognition between human decisions and recognition models. All correlations have p-values less than 1e-5.
}
\setlength\tabcolsep{5.0pt}
\footnotesize
\begin{tabular}{l*{5}{c}}
\toprule
 & \multicolumn{5}{c}{\textbf{Correlation Coefficient $\rho$ ($\uparrow$)}} \\ 
\cmidrule(lr){2-6}
\textbf{Method}                     & \textit{drums} & \textit{bass} & \textit{guitar} & \textit{piano} & \textit{avg.} \\ 
\midrule
\textbf{Success Rate} & 0.78            & 0.48           & 0.57             & 0.45            & 0.58           \\
\textbf{YAMNet}      & 0.80            & 0.37           & 0.56             & 0.33            & 0.57           \\
\textbf{PaSST}       & 0.79            & 0.49           & 0.59             & 0.43            & 0.57           \\
\textbf{CLAP}        & 0.80            & 0.53           & 0.56             & 0.52            & 0.65           \\
\textbf{ConvNeXt}    & 0.69            & 0.52           & 0.51             & 0.46            & 0.57           \\
\bottomrule
\end{tabular}
\vspace{-3mm}
\label{tab:subjective_inst_correlation}
\end{table}

\subsection{Subjective Evaluation}
\label{subsec:eval_sbj}

To further validate our intervention techniques on MusicGen, we performed a subjective listening test through Amazon Mechanical Turk following the best practices from~\cite{ribeiro2011crowdmos}. The purpose of this test was not to evaluate the quality and relevance of MusicGen as this was already explored in~\cite{musicgen}, but rather to provide evidence that: (1) our proposed \textit{success rate} objective metric correlates with human perception, and (2) our proposed intervention techniques are not detrimental to overall audio quality. To do this, we assessed whether humans could detect the presence of the instruments added by our intervention technique, and asked them to rate samples in terms of overall quality on a scale of 1-5. We also included real music samples from MoisesDB to provide a performance baseline.
In selecting files for each algorithm and instrument, we randomly selected 20 files above the median objective success rate and 20 files below the median. This leads to 1280 generated files to be rated (20 files x 2 top/bottom x 4 instruments x 4 algorithms x 2 audio continuation/text-to-music). All files in the listening test were ten seconds long and normalized to -12 loudness units full scale (LUFS)~\cite{grimm2010lufs}. We obtain at least 3 ratings per file. For the audio continuation experiments, we ask raters to ignore the first 3 seconds of each audio file, as that conditioning signal intentionally does not include the target instrument.

Table~\ref{tab:subjective_small} displays the overall objective quality results. %
In general, the performance of all algorithms is quite similar, with SMITIN and no intervention having the best overall quality.
In addition to overall quality, we also ask whether listeners can detect the presence of the instruments targeted by our intervention. This serves to help validate our success rate objective metric. We also compare the average human score with the average success rate objective metric (Alg.) for the top-ranked and bottom-ranked files selected for the listening test on each generation task in Table~\ref{tab:subjective_presence}. For the top-ranked files the scores appear to match quite well, while for the bottom-ranked files, it seems human raters tend to overestimate the presence of most instruments compared to the objective metric. We also hypothesize that the lower match for audio continuation compared to text-to-music may be due in part to the test being more subjectively difficult as raters have to focus on the end of the audio file under test.

Table~\ref{tab:subjective_inst_correlation} compares the Spearman correlation coefficient $\rho$ computed between each of the instrument recognition systems discussed in Section~\ref{subsubsec:obj_metric}) and the average human rating computed across all processing algorithms. Notably, using CLAP embeddings yielded the highest $\rho$ value with human decisions, while the other methods produced similar values. This result further supports the validity of using the Success Rate on generated samples. Not only does the Success Rate show a strong correlation with other recognition models, but the high correlation observed between the Success Rate and CLAP in Table~\ref{tab:obj_correlation} strengthens its reliability as an objective metric for controlling and evaluating generative music.

\section{Ablations and Practical Considerations}
\label{sec:analysis}

\begin{table}[t]
\centering
\sisetup{
    detect-weight,
    mode=text,
    tight-spacing=true,
    round-mode=places,
    round-precision=3,
    table-format=1.3,
    table-number-alignment=center
}
\vspace{-.24cm}
\caption{
    Ablation study on multi-directional ITI (performance averaged over all instruments)
}
\footnotesize
{
\resizebox{.9\linewidth}{!}
{
\begin{tabular}{lS[round-precision=1,table-format=2.1]S[round-precision=1,table-format=1.2]S[table-format=1.3]S[table-format=1.3]}
\toprule
 & 
\multicolumn{2}{c}{\textbf{Success Rate [\%]}} & 
 & 
 \\
\cmidrule(lr){2-3}
\textbf{Method}& {\textbf{indiv.}} & {\textbf{simult.}} &\textbf{FAD$_\text{L-CLAP mus}$} & \textbf{Similarity}\\
\midrule
\textbf{unconditioned}  & 11.1 & 0.1  & 0.419 & 0.781 \\
\textbf{text}     & 27.8 & 0.43 & 0.369 & 0.664 \\
\textbf{SMITIN}      & 25.3 & 1.23 & 0.445 & 0.734 \\
\bottomrule
\end{tabular}
}
}
\label{tab:ablation_multi_dir}
\vspace{-.4cm}
\end{table}

\noindent\textbf{Intervention with Multiple Directions.} We explore %
ITI in scenarios where multiple musical elements are introduced simultaneously. %
We focus on the continuation of music that has one instrument present and assesses the addition of three others. For instance, the intervention task generates a continuation with drums, bass, and piano when the input music only has guitar. We consider two types of success rates, an individual one for each instrument as before, and a simultaneous one to assess how often all 3 target instruments were added together. We consider a successful case for a song if each instrument is deemed present more than 50\% of the time.

The summarized results in Table~\ref{tab:ablation_multi_dir} reveal that while text prompts yield higher success rates for individual instruments, their generated output diverges from the input music's characteristics. This indicates that text conditioning may not adequately consider the input and opts to generate new content based solely on the given instruction. Without ITI, MusicGen tends to produce continuations focused on a single instrument, which deviates from the desired complex mixtures, as evidenced by the FAD scores. SMITIN, however, significantly outperforms text prompts in preserving input music similarity, and further enables more granular control over each musical aspect, leading to a higher success rate of jointly generating all desired instruments. This fine-grained control mechanism bolsters SMITIN's potential as a robust tool for complex music generation tasks where maintaining the essence of the input is crucial.

\noindent\textbf{Generating ``Realistic" Music.} To generate more realistic music through ITI, we fit probes to discern between real music (DISCO-10M dataset's DISCO-200K-high-quality subset \cite{discox}) and synthetic outputs from MusicGen itself. %
Despite MusicGen being trained only on real music,
the probing showed high performance with the best-performing probe achieving a 96.2\% accuracy rate and an average accuracy of 77.5\% in distinguishing real from synthetic music.
Leveraging this result, we steer MusicGen in an audio continuation task, using real-world music inputs from the DISCO-10M dataset to generate continuations that maintain the ``realistic" quality.

\begin{figure}[t]
     \centering
     \includegraphics[width=0.45\textwidth]{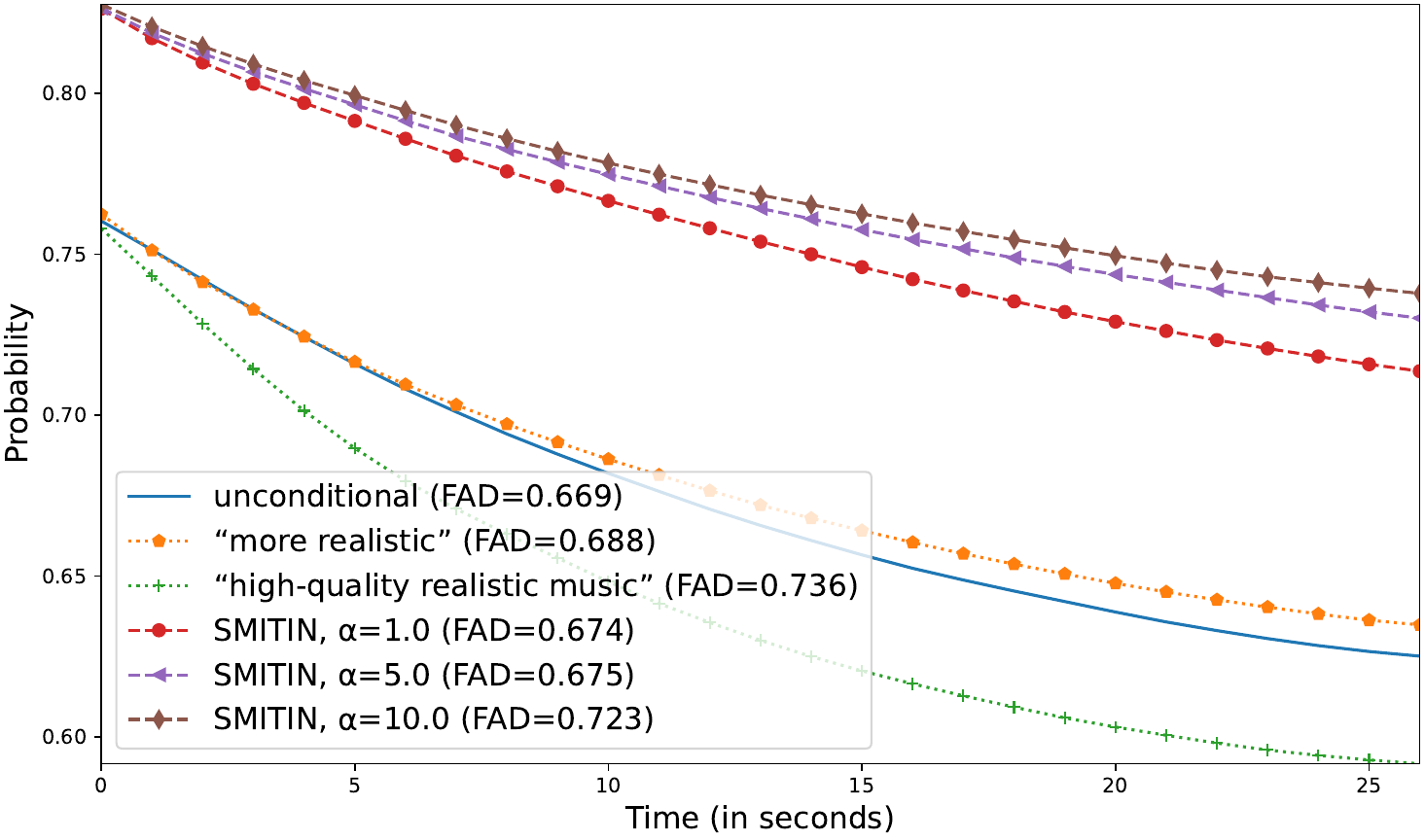}
     \caption{Temporal dynamics of maintaining ``realism" in audio continuation of real music sample. The graph tracks the probability of generated music being classified as $\langle$real$\rangle$ over time, where all configurations of SMITIN demonstrate an enhanced capacity to preserve realistic music qualities over time.}
    \label{fig:to_real}
    \vspace{-.8cm} %
\end{figure}

Analyzing the quantitative results depicted in Fig.~\ref{fig:to_real}, simple text prompts prove insufficient in preserving the realistic nature of the music; however, ITI emerges as a more effective approach. Regarding FAD scores, configurations of SMITIN with $\alpha$ values of 1 and 5 yield smaller deviations in distribution compared to text-based methods. Even with an increased intervention strength ($\alpha=10$), SMITIN does not shift the distribution more than that of the text prompt ``high-quality realistic music.''

\begin{table}[t]
\centering
\sisetup{
    detect-weight, %
    mode=text, %
    tight-spacing=true,
    round-mode=places,
    round-precision=3,
    table-format=1.3,
    table-number-alignment=center
}
\vspace{-.10cm}
\caption{
    Ablation on amount of probing data (performance is averaged over all instruments, $^*$shows full-set success rate).
}
{
\resizebox{1.\linewidth}{!}{
\begin{tabular}{lS[round-precision=1,table-format=2.1]*{3}{S}} %
\toprule
{\textbf{\shortstack[l]{Number of\\Paired Data}}} & 
\multicolumn{1}{c}{{\textbf{\shortstack[t]{Success\\Rate$^*$ [\%]}}}} & 
\multicolumn{1}{c}{\textbf{FAD$_\text{L-CLAP mus}$}} & 
{\textbf{Similarity}} & 
{\textbf{\shortstack[l]{Probes Max.\\Acc. [\%]}}} \\ %
\midrule
\textbf{$n$ = 10}    & 31.0 & 0.357 & 0.909 & {79.1} \\ %
\textbf{$n$ = 100}   & 31.8 & 0.349 & 0.914 & {82.6} \\
\textbf{$n$ = 500}   & 27.3 & 0.341 & 0.915 & {84.4} \\
\textbf{$n$ = 1000}  & 25.5 & 0.340 & 0.914 & {84.8} \\
\cdashlinelr{1-5}
\textbf{$n$ = full}  & 23.3 & 0.336 & 0.913 & {85.1} \\
\bottomrule
\end{tabular}
}
}
\label{tab:ablation_numdata}
\vspace{-.5cm}
\end{table}

\noindent\textbf{Effects of Number of Probing Data.} To investigate the influence of probing data quantity on SMITIN, we test the system with varying sizes of probing datasets randomly selected from the complete set. These subsets consist of $n=10$, $100$, $500$, and $1000$ data pairs, equivalent to music durations of $1$, $10$, $50$, and $100$ minutes respectively. 
The results are presented in Table~\ref{tab:ablation_numdata}, where we show the ``full-set success rate'' which is computed using the classifier probes trained with the full dataset.%

The results indicate that the system demonstrates reasonable results even with a minimal dataset of just 1 minute. As the data size increases to $n=500$ (50 minutes), the objective metrics stabilize, indicating an optimal data quantity threshold for effective probe training. An interesting observation from our results is the higher full-set success rates associated with smaller datasets. This trend suggests that probes trained on limited data may have lower confidence in their decisions, prompting them to recommend more interventions during the ITI process to meet the success threshold. At the same time, we observe a clear correlation between better probes (represented here through the accuracy of the best probe) and better ITI performance, whether measured as a higher full-set success rate or similarity, or lower FAD. 

This suggests that users can effectively leverage ITI with only a small amount of data to control generation. It means that effective ITI for controlling any musical trait can be accessible and achievable without the need for extensive datasets, which is particularly beneficial for individual artists or smaller studios. This broadens the potential for creative and personalized applications of ITI in music generation, making it a versatile tool for a wide range of users.

\vspace{-.3cm}
\section{Conclusions and Future Work}
\label{sec:conclusion}
We proposed a novel approach for inference-time control of generative music transformers, which self-monitors probe accuracy to impose desired musical traits while maintaining overall music quality.  A limitation of our approach is the dependence of performance on probe accuracy, that is, if the pre-trained transformer has not accurately learned a concept, or if the probe training set is inadequate, the success rate of our intervention technique may suffer (although it still could be creatively useful). In the future, we plan to explore different generative model architectures~\cite{garcia2023vampnet, evans2024stable, hewitt2023backpack} and interventions using musical traits such as genre, emotion, etc. Furthermore, we hope to investigate building ``knobs'' for interactively controlling the hyper-parameters necessary for inference-time control to enable new technical and creative interactions.

\balance
\bibliographystyle{IEEEtran}
\bibliography{reference}

\section*{Appendix}
\subsection{Analytical Insights from Probing}
\label{subsec:probing_analysis}

\begin{table*}[t]
\centering
\caption{Comparison of multi-label instrument recognition for head-wise probing on MusicGen and a supervised model.}
\resizebox{0.87\linewidth}{!}{
\begin{tabular}{lcccccc}
\toprule
 & \multicolumn{5}{c}{\textbf{Accuracy / F1 Score}} & \multirow{2}{*}{\textbf{Num.}} \\
\cmidrule(lr){2-6}
\textbf{Model} & \textit{vocals} & \textit{bass} & \textit{drums} & \textit{other} & \textit{avg.} & \textbf{Param.} \\
\midrule
ConvNeXt$_\text{tiny}$\cite{liu2022convnet} & \textbf{97.8\%} / \textbf{0.947} & \textbf{94.4\%} / 0.891 & 95.1\% / 0.914 & \textbf{93.2\%} / \textbf{0.880} & \textbf{95.1\%} / 0.906 & 28.5M \\ \midrule %
MusicGen$_\text{small}$ & 92.5\% / 0.929 & 91.6\% / 0.920 & 95.0\% / 0.952 & 87.1\% / 0.868 & 91.5\% / 0.917 & 300M \\
MusicGen$_\text{medium}$ & 92.9\% / 0.934 & 91.8\% / \textbf{0.923} & 94.4\% / 0.947 & 87.4\% / 0.874 & 91.6\% / 0.919 & 1.5B \\
MusicGen$_\text{large}$ & 92.8\% / 0.932 & 91.7\% / 0.920 & 95.1\% / 0.953 & 85.7\% / 0.872 & 91.8\% / 0.919 & 3.3B \\
MusicGen$_\text{melody}$ & 94.1\% / 0.945 & 91.8\% / \textbf{0.923} & \textbf{95.8\%} / \textbf{0.960} & 87.9\% / 0.876 & 92.4\% / \textbf{0.926} & 1.5B \\
\bottomrule
\end{tabular}
}
\label{tab:multi_inst_recognition}
\end{table*}

Following the probing method outlined in the main paper, we investigate additional downstream tasks through probing to present more objective results and enhance our understanding. First, we assess MusicGen's probing capabilities by comparing them to ConvNeXt~\cite{liu2022convnet}, a model trained via supervised learning. Subsequently, we contrast our results with prior work on music probing~\cite{jukebox_probing} which uses entire attention layer outputs from Jukebox, another generative music transformer model.

\subsubsection{Multi-label Instrument Recognition}
\label{subsubsec:multi_inst_recognition}

Following up on the performance of single-instrument recognition, we compare the capabilities of individual attention heads in MusicGen against a supervised model trained for the multi-label instrument recognition task, i.e,  identifying the presence of all instruments of a given music clip. This is achieved by fitting multiple logistic regression classifiers (probes) for each instrument class using again the extracted intermediate MusicGen activations at the last time step. For the final evaluation, we select the best-performing probe for each instrument class, noting that these may originate from different attention layers ($l$) and heads ($h$). Following the methodology described in \cite{koo2023self}, we conduct recognition tasks on 3-second music segments from the MUSDB dataset \cite{musdb18-hq}. We also report the performance of a model trained in a supervised manner specifically for instrument recognition~\cite{koo2023self}, utilizing the `\textit{tiny}' configuration of ConvNeXt \cite{liu2022convnet}.

The objective results for multi-label instrument recognition are detailed in Table \ref{tab:multi_inst_recognition}. We find that the performance of MusicGen's probes is comparable to that of the supervised method.
However, across all configurations of MusicGen, we note strong performance in terms of the F1 score. Specifically, MusicGen exhibits exceptional performance in recognizing \textit{drums}, whereas its least effective performance is observed in the \textit{other} category. This discrepancy likely arises from the ambiguous definition of the \textit{other} category in the MUSDB18 dataset, with the supervised method gaining an advantage by explicitly training on the dataset's label, thereby enhancing its ability to identify \textit{other}.

\begin{table*}[t]
\centering
    \sisetup{
    detect-weight, %
    mode=text, %
    tight-spacing=true,
    round-mode=places,
    round-precision=1,
    table-format=2.1,
    table-align-text-post=false,
    table-number-alignment=center
    }
\caption{
    Performance comparison of layer-wise and head-wise probing on generative music transformers. For music tagging, we report the performance of the best-performing probe across all classes, with the ensemble result of the best-performing probes for each class indicated within parentheses.
}
\begin{tabular}{l*{7}{S}cc}
\toprule
\multicolumn{1}{r}{\textbf{Dataset}} & \multicolumn{2}{c}{\textbf{MTT}} & \textbf{GTZAN} & \textbf{GS}          & \multicolumn{2}{c}{\textbf{EMO}} &  & &  \\
\multicolumn{1}{r}{\textbf{Task}}    & \multicolumn{2}{c}{Tagging}      & {Genre}          & {Key}                  & \multicolumn{2}{c}{Emotion}      &                                &                                      &     \multirow[b]{2.55}{*}{\shortstack[t]{\textbf{Num.}\\\textbf{Param.}}   }        \\ 
\cmidrule(lr){2-3}\cmidrule(lr){4-4}\cmidrule(lr){5-5}\cmidrule(lr){6-7}
\multicolumn{1}{r}{\textbf{Metrics}} & {AUC}             & {AP}             & {Acc}            & {Acc$^\text{ref.}$} & {R2$^A$}          & {R2$^V$}         &    {{Avg.}}     & {\textbf{Dim.}}    &                                          \\ \midrule
Jukebox                              & 91.5            & 41.4           & 79.7           & 66.7                 & 72.1            & 61.7           & 69.9                           & 4.8K                                 & 5B                                                                              \\ \midrule \midrule
MusicGen$_\text{small}$              &  \multicolumn{1}{c}{85.5 (86.7)}            & \multicolumn{1}{c}{34.1 (37.5)}           & 66.2           & 46.6                 & 64.2            & 43.5           & 55.2                           & 64                                   & 300M                                                                            \\
MusicGen$_\text{medium}$             & \multicolumn{1}{c}{85.9 (87.3)}            & \multicolumn{1}{c}{33.9 (38.4)}           & 69.7           & 57.4                 & 65.3            & 51.6           & 59.6                           & 64                                   & 1.5B                                                                            \\
MusicGen$_\text{large}$              & \multicolumn{1}{c}{85.1 (87.2)}            & \multicolumn{1}{c}{32.9 (38.5)}           & 71.0           & 58.5                 & 69.1            & 49.3           & 60.3                           & 64                                   & 3.3B                                                                            \\
MusicGen$_\text{melody}$             & \multicolumn{1}{c}{85.8 (87.1)}            & \multicolumn{1}{c}{33.3 (38.1)}           & 65.2           & 62.1                 & 64.7            & 44.8           & 58.8                           & 64                                   & 1.5B                                                                            \\ \bottomrule
\end{tabular}
\vspace{-10pt}
\label{tab:probing_downstream_tasks}
\end{table*}
\begin{figure*}[t]
     \centering
     \subfigure{\includegraphics[width=0.188\textwidth]{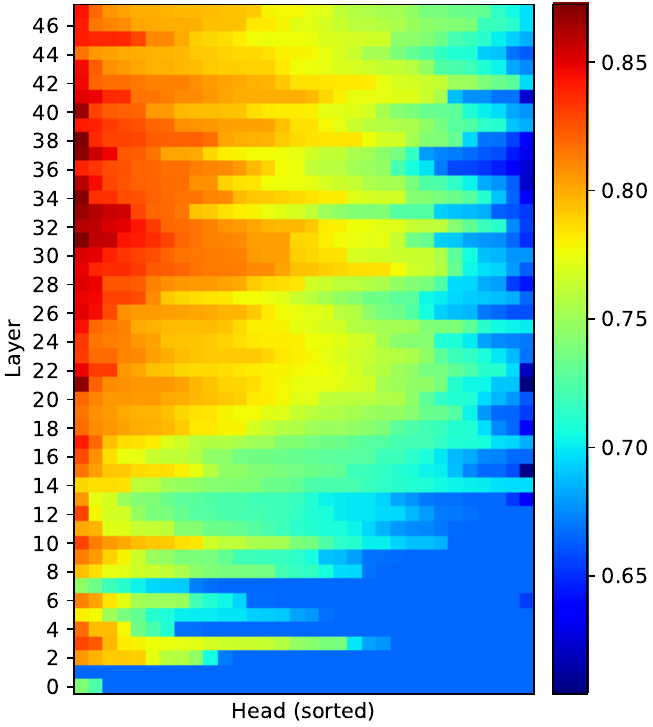}
     \label{subfig:probe_acc_multi_inst}}
     \hfill
     \subfigure{\includegraphics[width=0.188\textwidth]{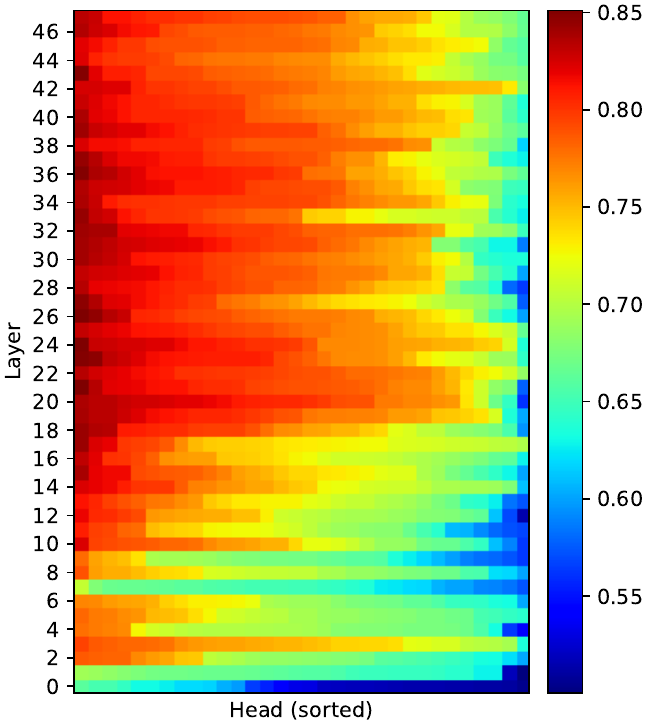}
     \label{subfig:probe_acc_tagging}}
     \hfill
     \subfigure{\includegraphics[width=0.188\textwidth]{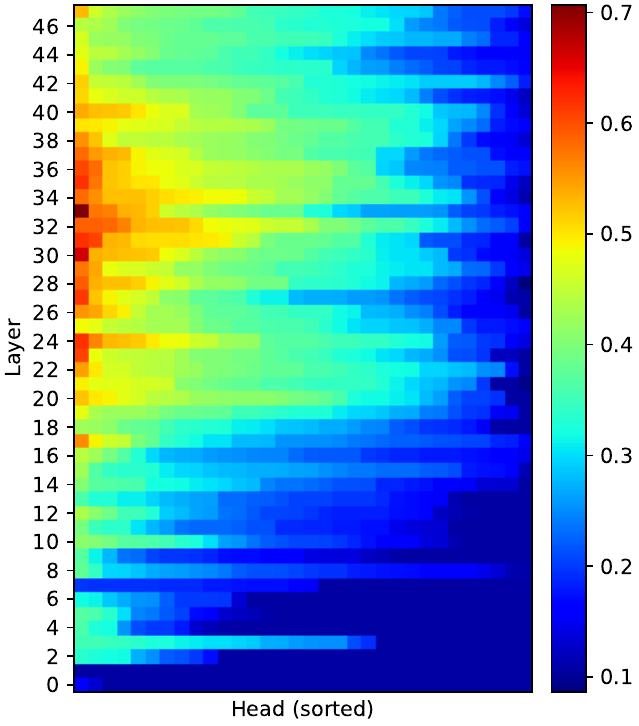}
     \label{subfig:probe_acc_genre}}
     \hfill
     \subfigure{\includegraphics[width=0.188\textwidth]{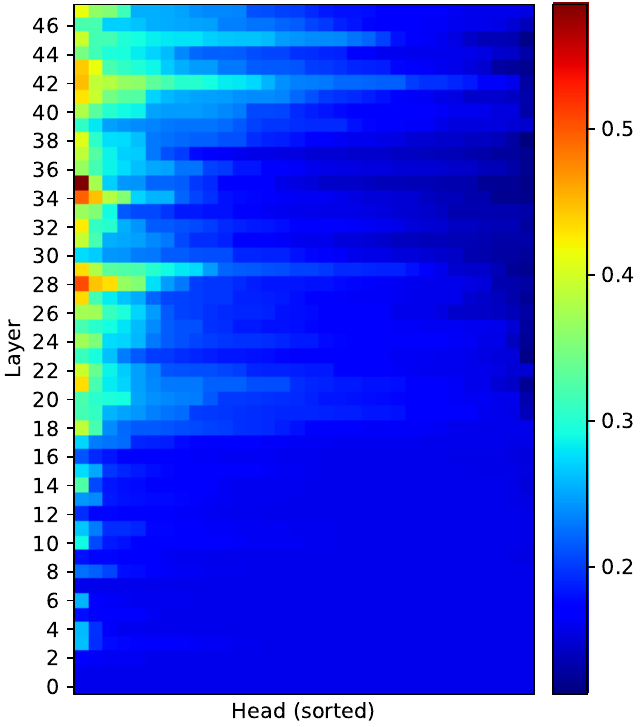}
     \label{subfig:probe_acc_key_detection}}
     \hfill
     \subfigure{\includegraphics[width=0.188\textwidth]{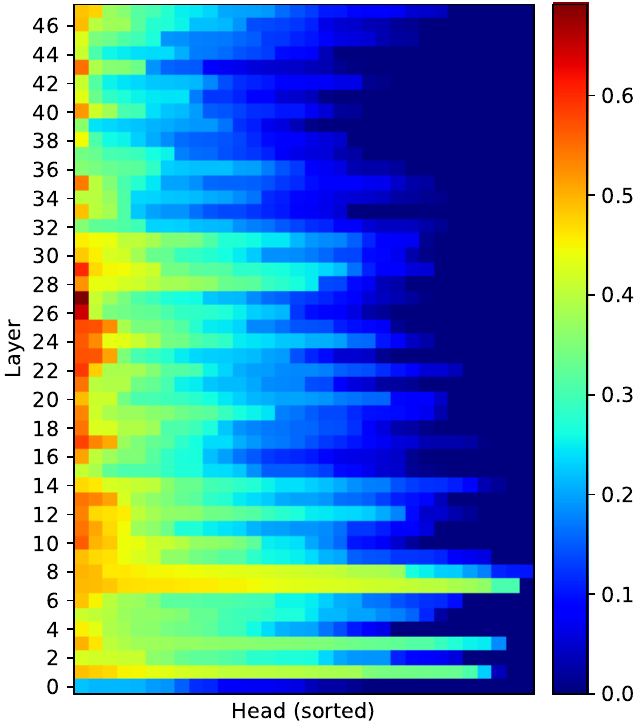}
     \label{subfig:probe_acc_emotion}}
     \setcounter{subfigure}{0}
     \subfigure[Multi-label Inst.\ (F1)]{\includegraphics[width=0.188\textwidth]{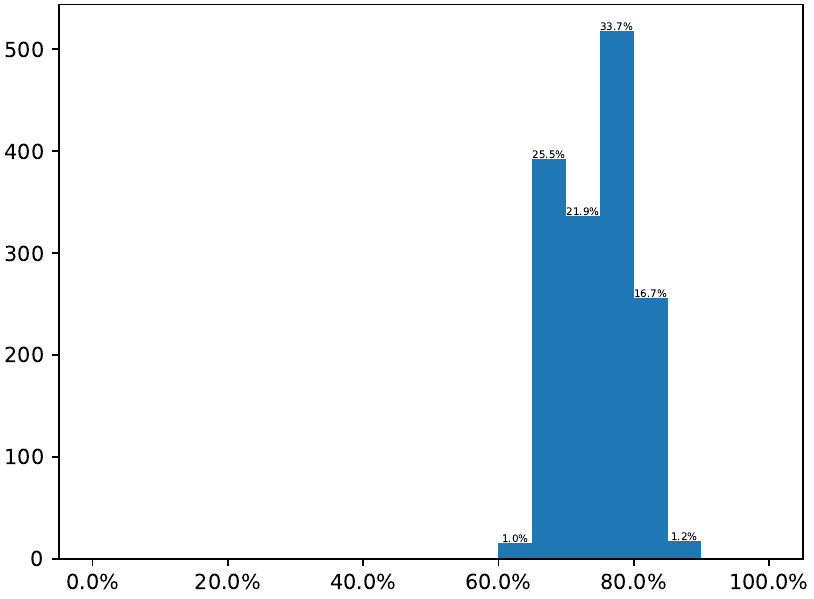}
     \label{subfig:probe_hist_multi_inst}}
     \hfill
     \subfigure[Tagging (AUC)]{\includegraphics[width=0.188\textwidth]{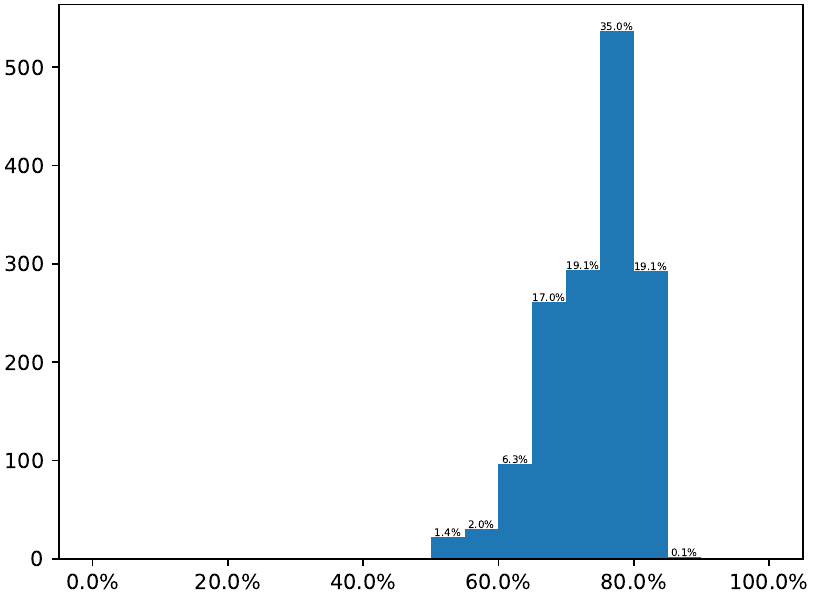}
     \label{subfig:probe_hist_tagging}}
     \hfill
     \subfigure[Genre (Acc)]{\includegraphics[width=0.188\textwidth]{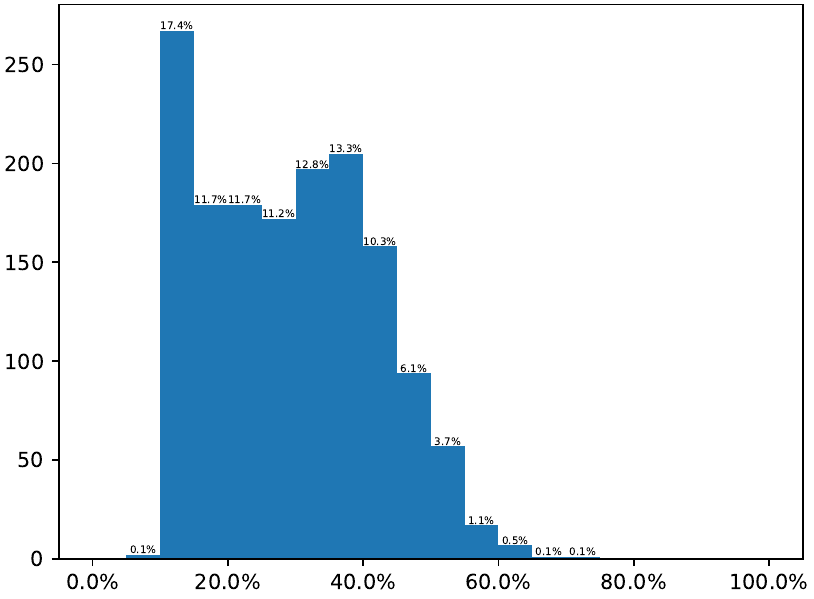}
     \label{subfig:probe_hist_genre}}
     \hfill
     \subfigure[Key (Acc$^\text{ref.}$)]{\includegraphics[width=0.188\textwidth]{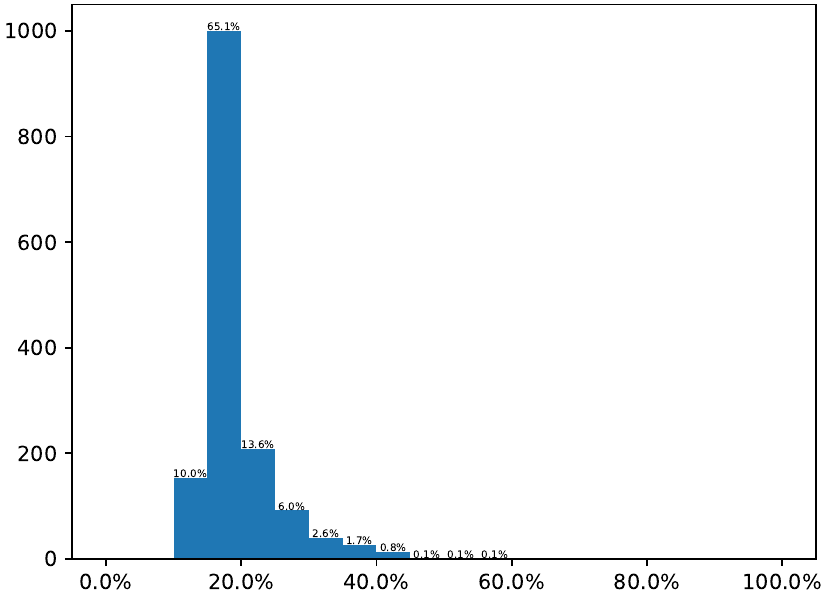}
     \label{subfig:probe_hist_key_detection}}
     \hfill
     \subfigure[Emotion (R2$^A$)]{\includegraphics[width=0.188\textwidth]{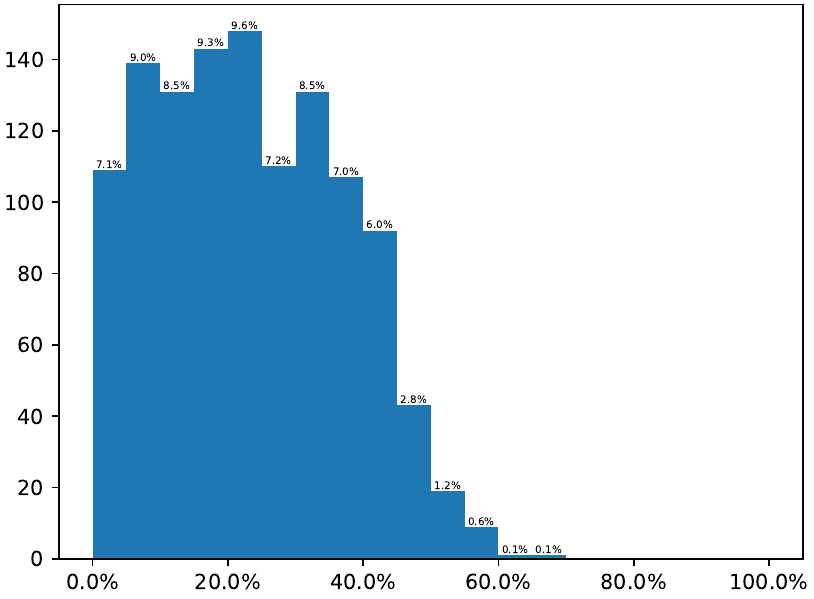}
     \label{subfig:probe_hist_emotion}}
     \caption{Probe performance on various music understanding tasks. Each set of figures shows the performance on each task with probes fitted with MusicGen$_\text{large}$ model output activations. Histograms in the bottom row show the distribution of head-wise probes according to their performance.}
    \label{fig:probe_acc_main}
    \vspace{-10pt}
\end{figure*}

\subsubsection{Various Music Downstream Tasks}
\label{subsubsec:music_downstream_tasks}

To expand our understanding of MusicGen's capabilities beyond instrument recognition, we conducted probing on a set of more general music understanding tasks. These include music tagging (MTT) \cite{mtat}, genre classification (GTZAN) \cite{gtzan}, key detection (GS) \cite{gs_key_detection}, and emotion recognition (EMO) \cite{emo_recognition}. We adopt appropriate regression models for probing for these tasks: multinomial logistic regression for multi-class classification tasks, multiple logistic regressions for multi-label classification tasks, and linear regression for regression tasks. For these tasks, we set the input duration as 29.0 seconds, and let the probes take the last time step activation output for classification.

Table \ref{tab:probing_downstream_tasks} presents the probing results for various configurations of MusicGen using our attention head probing, and a prior music probing work \cite{jukebox_probing} that relies on a comparative model, Jukebox \cite{jukebox}, but uses a different probing methodology:
\begin{enumerate}[leftmargin=*]
    \item \cite{jukebox_probing} probed Jukebox using entire intermediate attention layers, whereas our approach evaluates performance based on individual attention heads.
    
    \item \cite{jukebox_probing} employed a one-layer MLP with 512 hidden units. In preliminary experiments, using a one-layer MLP for head-wise probes showed limited benefits, possibly because the attention dimension per head that we use is 75 times smaller than that of Jukebox.
\end{enumerate}
Jukebox probing typically shows superior performance, maybe due to the model's larger number of parameters and activation dimension. In music tagging, we present both the performance of the best-performing probe across all classes and the ensemble result of the best-performing probes for each class, indicated within parentheses. We observe an improved performance with the ensemble technique, though the improvement is typically not substantial.
Interestingly, this means that we can find a single head capable of understanding multiple attributes simultaneously, a notable feature given the MTT dataset comprises 188 different tags. While a larger number of parameters generally correlates with enhanced performance, the melody configuration outperforms others in key detection. This superior performance is likely attributed to the model's training with a melody conditioning objective~\cite{musicgen}.

Figure \ref{fig:probe_acc_main} provides a qualitative visualization of the performance of MusicGen$_\text{large}$ probes across various music understanding tasks, alongside histograms representing the distribution of head-wise performance. In tasks such as multi-label instrument recognition and music tagging, there is a noticeable similarity in the trends observed both in the histograms and the layer-wise performance distributions. For key detection, the results indicate that comprehension is limited to only a few heads. On the other hand, emotion recognition demonstrates a relatively uniform performance across the majority of heads in the earlier layers, particularly layer 7. However, the highest-performing head for emotion recognition is located in a middle layer, specifically at the 27$^\text{th}$ layer. This analysis underscores the nuanced role that different attention heads play in music understanding, highlighting the potential for precise control on each attention head to optimize performance for specific tasks within generative music models.

\subsection{Subjective Listening Test Setup and Detailed Results}
\label{appendix:listening_test}

In selecting files for the listening test discussed in the main paper, for each algorithm and instrument addition intervention, we randomly select 20 files above the median objective success rate, and 20 files below the median. This leads to 1280 generated files to be rated (20 files x 2 top/bottom x 4 instruments x 4 algorithms x 2 audio continuation/text-to music). We additionally include real music samples from MoisesDB as references. All audio files included in the listening test were ten seconds in length and normalized to -12 loudness units full scale (LUFS)~\cite{grimm2010lufs}. We obtain at least 3 ratings per file. For the audio continuation experiments, we ask raters to ignore the first 3 seconds of each audio file, as that conditioning signal intentionally does not include the target instrument. In addition to following the best practices from~\cite{ribeiro2011crowdmos}, we limited participation to ``Mechanical Turk Masters'' for which Amazon charged a \$0.005 fee per assignment, and paid \$0.10 per rated file.

Table~\ref{tab:subjective_mos_full} displays the overall objective quality results broken out by instrument and algorithm. In general, the performance of all algorithms are quite similar, with SMITIN an no intervention having the best overall quality.

In addition to overall quality, we also ask whether listeners can detect the presence of the instruments targeted by our intervention. This serves to help validate our success rate objective metric, which was shown and discussed in the main paper. 

\begin{table*}[t]
\centering
\sisetup{
detect-weight, %
mode=text, %
tight-spacing=true,
round-mode=places,
round-precision=2,
table-format=3.2,
table-number-alignment=center
}
\caption{Average and 95\% confidence interval of subjective listening tests for overall audio quality across four instrument addition interventions for both audio continuation and text-to-music experiments.
}
\label{tab:subjective_mos_full}
\setlength\tabcolsep{2.0pt}
\footnotesize
\resizebox{1.0\linewidth}{!}
    {
\begin{tabular}{l*{20}{S@{\,\( \pm \)\,}S[table-format=1.2]}}
\toprule
 & \multicolumn{10}{c}{\textbf{Audio continuation}} & \multicolumn{10}{c}{\textbf{Text-to-music}}  \\
\cmidrule(lr){2-11}\cmidrule(lr){12-21}
\textbf{Method} & \multicolumn{2}{c}{\textit{drums}}      & \multicolumn{2}{c}{\textit{bass}} & \multicolumn{2}{c}{\textit{guitar}} & \multicolumn{2}{c}{\textit{piano}}   & \multicolumn{2}{c}{\textit{avg}}  & \multicolumn{2}{c}{\textit{drums}}      & \multicolumn{2}{c}{\textit{bass}} & \multicolumn{2}{c}{\textit{guitar}} & \multicolumn{2}{c}{\textit{piano}}   & \multicolumn{2}{c}{\textit{avg}}   \\ \midrule
\textbf{No intervention}      &  3.22 & 0.21   & 3.75 & 0.18 &  3.26 & 0.22   & 3.16 & 0.18 &  3.35 & 0.10   & 3.63 & 0.18 &  3.66 & 0.20   & 3.70 & 0.19 &  3.70 & 0.17   & 3.67 & 0.09 \\
\textbf{``add \textless{}inst.\textgreater{}"}     &  3.31 & 0.21   & 3.41 & 0.18 &  3.06 & 0.211   & 3.31 & 0.22 &  3.27 & 0.10   & 3.78 & 0.18 &  3.61 & 0.18   & 3.54 & 0.21 &  3.50 & 0.19   & 3.61 & 0.09 \\
\textbf{original ITI}      &  3.42 & 0.22   & 3.04 & 0.21 &  3.54 & 0.20   & 3.12 & 0.22 &  3.28 & 0.11   & 3.62 & 0.19 &  3.61 & 0.18   & 3.65 & 0.19 &  3.33 & 0.20   & 3.55 & 0.09 \\
\textbf{SMITIN}      &  3.40 & 0.20   & 3.45 & 0.18 &  3.27 & 0.21   & 3.23 & 0.19 &  3.34 & 0.10   & 3.93 & 0.18 &  3.75 & 0.17   & 3.77 & 0.18 &  3.73 & 0.17   & 3.80 & 0.09 \\\midrule
\textit{MoisesDB}    &  \multicolumn{20}{c}{\textit{3.98 $\pm$ 0.29}} \\ \bottomrule
\end{tabular}
}
\end{table*}

\begin{table*}[t]
\centering
\sisetup{
    detect-weight,
    mode=text,
    tight-spacing=true,
    round-mode=places,
    round-precision=1,
    table-format=2.1,
    table-number-alignment=center
}
\caption{
    Intervention with Multiple Directions (Individual Success Rate).
}
\setlength\tabcolsep{3.5pt}
\footnotesize
\begin{tabular}{
    l
    S[table-format=2.1]
    S[table-format=2.1]
    S[table-format=1.1]
    S[table-format=2.1]
    S[table-format=2.1]
    S[table-format=2.1]
    S[table-format=1.1]
    S[table-format=2.1]
    S[table-format=2.1]
    S[table-format=2.1]
    S[table-format=1.1]
    S[table-format=2.1]
    S[table-format=2.1]
    S[table-format=2.1]
    S[table-format=2.1]
    S[table-format=2.1]
    S[table-format=2.1]
}
\toprule
 & \multicolumn{16}{c}{\textbf{Individual Success Rate [\%]}} & \\
\cmidrule(lr){2-17}
& \multicolumn{4}{c}{start with \textit{drums}} & \multicolumn{4}{c}{start with \textit{bass}} & \multicolumn{4}{c}{start with \textit{guitar}} & \multicolumn{4}{c}{start with \textit{piano}} \\
\cmidrule(lr){2-5}\cmidrule(lr){6-9}\cmidrule(lr){10-13}\cmidrule(lr){14-17}
\textbf{Method} & {\textit{bass}} & {\textit{guitar}} & {\textit{piano}} & {\textit{avg.}} & {\textit{drums}} & {\textit{guitar}} & {\textit{piano}} & {\textit{avg.}} & {\textit{drums}} & {\textit{bass}} & {\textit{piano}} & {\textit{avg.}} & {\textit{drums}} & {\textit{bass}} & {\textit{guitar}} & {\textit{avg.}} & \textit{total} \\
\midrule
\textbf{unconditioned} & 29.0 & 4.3 & 0.2 & 11.1 & 17.7 & 10.6 & 1.4 & 9.9 & 18.8 & 18.4 & 1.9 & 13.0 & 11.9 & 10.4 & 9.1 & 10.4 & 11.1 \\
\textbf{text} & 58.4 & 2.6 & 5.2 & 22.0 & 45.7 & 2.8 & 2.9 & 17.1 & 45.5 & 67.0 & 4.8 & 39.1 & 42.9 & 50.4 & 6.6 & 33.3 & 27.8 \\
\textbf{SMITIN} & 48.3 & 16.1 & 01.6 & 22.0 & 25.4 & 23.6 & 04.1 & 17.7 & 41.1 & 49.0 & 3.3 & 31.1 & 30.2 & 35.7 & 26.2 & 30.7 & 25.3 \\
\bottomrule
\end{tabular}
\label{tab:multi_dir_indiv}
\end{table*}

\begin{table*}[t]
\centering
\sisetup{
    detect-weight,
    mode=text,
    tight-spacing=true,
    round-mode=places,
    round-precision=3,
    table-format=1.3,
    table-number-alignment=center
    }
\caption{
    Intervention with Multiple Directions (Simultaneous Success Rate, FAD, and Similarity).
}
\setlength\tabcolsep{3.5pt}
\footnotesize
\begin{tabular}{
    l
    S[round-precision=1,table-format=1.1]
    S[round-precision=1,table-format=1.1]
    S[round-precision=1,table-format=1.1]
    S[round-precision=1,table-format=1.1]
    S[round-precision=1,table-format=1.1]
    c*{10}{S}
}
\toprule

& \multicolumn{5}{c}{\textbf{Simultaneous Success Rate [\%] ($\uparrow$)}}                                       & \multicolumn{5}{c}{\textbf{FAD$_\text{L-CLAP mus}$ ($\downarrow$)}}  &  \multicolumn{5}{c}{\textbf{Similarity ($\uparrow$)}}   \\ 
\cmidrule(lr){2-6}\cmidrule(lr){7-11}\cmidrule(lr){12-16}
\textbf{Method} & \textit{drums} & \textit{bass} & \textit{guitar} & \textit{piano} & \textit{avg.} & \textit{drums} & \textit{bass} & \textit{guitar} & \textit{piano} & \textit{avg.} & \textit{drums} & \textit{bass} & \textit{guitar} & \textit{piano} & \textit{avg.} \\ 
\midrule
\textbf{unconditioned} & 0.0 & 0.0 & 0.2 & 0.2 & 0.1 &
0.442 & 0.517 & 0.280 & 0.440 & 0.419 &
0.884 & 0.580 & 0.844 & 0.817 & 0.781\\
\textbf{text} & 0.3 & 0.0 & 0.5 & 0.9 & 0.42 &
0.364 & 0.426 & 0.376 & 0.312 & 0.369 &
0.847 & 0.565 & 0.589 & 0.655 & 0.664 \\
\textbf{SMITIN} & 0.1 & 0.4 & 0.5 & 3.9 & 1.225 &
0.467 & 0.562 & 0.3 & 0.452 & 0.445 &
0.837 & 0.549 & 0.79 & 0.763 & 0.734 \\
\bottomrule
\end{tabular}
\label{tab:multi_dir_full}
\end{table*}

\begin{table*}[t]
\centering
\sisetup{
    detect-weight,
    mode=text,
    tight-spacing=true,
    round-mode=places,
    round-precision=3,
    table-format=1.3,
    table-number-alignment=center
    }
\caption{
    Objective evaluation on \textit{Instrument Removal}.
}
\setlength\tabcolsep{3.5pt}
\footnotesize
\begin{tabular}{cl*{5}{S[round-precision=1,
    table-format=2.1]}*{5}{S}*{5}{S}}
\toprule
\multicolumn{2}{c}{\textbf{Method}}
& \multicolumn{5}{c}{\textbf{Success Rate [\%] ($\uparrow$)}}                                       & \multicolumn{5}{c}{\textbf{FAD$_\text{L-CLAP mus}$ ($\downarrow$)}}  &  \multicolumn{5}{c}{\textbf{Similarity ($\uparrow$)}}   \\ 
\cmidrule(lr){1-2}\cmidrule(lr){3-7}\cmidrule(lr){8-12}\cmidrule(lr){13-17}
\textbf{ITI}                & \textbf{Configuration}                   & \textit{drums} & \textit{bass} & \textit{guitar} & \textit{piano} & \textit{avg.} & \textit{drums} & \textit{bass} & \textit{guitar} & \textit{piano} & \textit{avg.} & \textit{drums} & \textit{bass} & \textit{guitar} & \textit{piano} & \textit{avg.}  \\ 
\midrule
\multirow{3}{*}{\textit{\sffamily X}}  & \textbf{unconditioned} & 2.2 & 16.2 & 14.1 & 31.6 & 16.0 &
0.442 & 0.517 & 0.338 & 0.513 & 0.452 &
0.884 & 0.579 & 0.847 & 0.816 & 0.781 \\
& \textit{\textbf{``remove \textless{}inst.\textgreater{}''}} & 
1.8 & 4.8 & 19.2 & 23.2 & 12.2 &
0.568 & 0.634 & 0.420 & 0.789 & 0.602 &
0.871 & 0.545 & 0.795 & 0.808 & 0.754
\\ 
& \textit{\textbf{``no \textless{}inst.\textgreater{}''}} &
7.3 & 10.8 & 17.0 & 32.5 & 16.9 &
0.267 & 0.267 & 0.307 & 0.289 & 0.282 & 
0.934 & 0.889 & 0.937 & 0.943 & 0.925
\\
\cdashlinelr{1-17}
\multirow{2}{*}{\textbf{\checkmark}} & \textbf{original ITI} & 
11.5 & 0.04 & 84.9 & 95.3 & 47.9 &
0.508 & 0.598 & 0.686 & 0.565 & 0.589 & 
0.833 & 0.835 & 0.702 & 0.867 & 0.809
\\
& \textbf{SMITIN} & 
9.1 & 5.1 & 79.9 & 78.4 & 43.1 &
0.481 & 0.538 & 0.481 & 0.455 & 0.488 & 
0.779 & 0.545 & 0.691 & 0.778 & 0.698
\\
\bottomrule
\end{tabular}
\label{tab:obj_eval_inst_removal}
\end{table*}

\subsection{Intervention with Multiple Directions}
\label{subsec:multiple_direction}

We explore the application of ITI in scenarios where multiple musical elements are introduced simultaneously. In our setup, we focus on the continuation of music that has one instrument present and assesses the addition of three others. For instance, the intervention task is to generate a continuation with drums, bass, and piano when the input music only has guitar. We consider two types of success rates, an individual one for each instrument as before, and a simultaneous one to assess how often all 3 target instruments were added together. We consider a successful case for a song if each instrument is deemed present more than 50\% of the time.

The summarized results in Tables~\ref{tab:multi_dir_indiv} and~\ref{tab:multi_dir_full} reveal that while text prompts yield higher success rates for individual instruments, their generated output diverges from the input music's characteristics. This indicates that text conditioning may not adequately consider the input and opts to generate new content based solely on the given instruction. Without ITI, MusicGen tends to produce continuations focused on a single instrument, which deviates from the desired complex mixtures, as evidenced by the FAD scores. SMITIN, however, significantly outperforms text prompts in preserving input music similarity, and further enables more granular control over each musical aspect, leading to a higher success rate of jointly generating all desired instruments. This fine-grained control mechanism bolsters SMITIN's potential as a robust tool for complex music generation tasks where maintaining the essence of the input is crucial.

\subsection{Instrument Removal}
\label{subsec:appendix_inst_removal}

We explore an additional application: \textit{removing} the target instrument during audio continuation using the same probes employed for instrument addition. To accomplish this, we apply negative intervention strength $\alpha$ to eliminate the target instrument from the given input music. The hyperparameter settings for this task are configured as $\alpha=-10.0$ for both original ITI and SMITIN, with $s=1$ and $\tau=0.5$ for SMITIN. The success rate in this context is defined as $\bar{\mathcal{C}}(t) < \tau$, where $\tau=0.5$.

Table~\ref{tab:obj_eval_inst_removal} reveals an inverse tendency for each instrument's removal: drums and bass tend to be continuously generated, whereas guitar and piano are more likely to be omitted during the generation process. Moreover, we find that both text-conditioning prompts: \textit{``remove \textless{}inst.\textgreater{}"}, and \textit{``no \textless{}inst.\textgreater{}"} are ineffective in removing the target instrument. This outcome suggests another aspect of controllability that is challenging to achieve with text prompts alone.

\begin{table*}[ht]
\centering
\sisetup{
    detect-weight, %
    mode=text, %
    tight-spacing=true,
    round-mode=places,
    round-precision=3,
    table-format=1.3,
    table-number-alignment=center
    }
\caption{
    Ablation study: top-$K$ head selection vs.\ head soft-weighting
}
\setlength\tabcolsep{3.5pt}
\footnotesize
{
\begin{tabular}{c*{5}{S[round-precision=1,
    table-format=2.1]}*{5}{S}*{5}{S}}
\toprule
\multirow[b]{2.5}{*}{\textbf{\shortstack{Top-K\\Interventions}}} 
& \multicolumn{5}{c}{\textbf{Success Rate [\%] ($\uparrow$)}}                                       & \multicolumn{5}{c}{\textbf{FAD$_\text{L-CLAP mus}$ ($\downarrow$)}}  &  \multicolumn{5}{c}{\textbf{Similarity ($\uparrow$)}}   \\ 
\cmidrule(lr){2-6}\cmidrule(lr){7-11}\cmidrule(lr){12-16}
 & \textit{drums} & \textit{bass} & \textit{guitar} & \textit{piano} & \textit{avg.} & \textit{drums} & \textit{bass} & \textit{guitar} & \textit{piano} & \textit{avg.} & \textit{drums} & \textit{bass} & \textit{guitar} & \textit{piano} & \textit{avg.} \\ 
\midrule
\textbf{K = 16\phantom{00}}                                                                           & 14.9          & 24.8          & 11.9           & 3.9          & 13.8          & 0.327          & \textbf{0.249} & \textbf{0.392}  & 0.335          & \textbf{0.325}   & 0.864          & \textbf{0.929} & 0.931           & 0.941          & 0.916          \\
\textbf{K = 32\phantom{00}}                                                                           & 14.4          & 20.2          & 16.9           & 5.6          & 14.2          & \textbf{0.325} & 0.259          & 0.393           & 0.334          & 0.327         & 0.867          & 0.928          & 0.934           & 0.941          & \textbf{0.917} \\
\textbf{K = 64\phantom{00}}                                                                           & 15.3          & 26.5          & 17.4           & 11.0          & 17.5          & 0.335          & 0.267          & 0.390           & 0.336          & 0.332       & \textbf{0.869} & 0.923          & 0.934           & 0.942          & \textbf{0.917} \\
\textbf{K = 128\phantom{0}}                                                                          & 18.2          & 33.6          & 27.4           & 13.8          & 23.2          & 0.352          & 0.277          & 0.411           & \textbf{0.327} & 0.341          & 0.860          & 0.924          & 0.932           & \textbf{0.944} & 0.915          \\
\textbf{K = 256\phantom{0}}                                                                        & 23.9          & 36.0          & 36.8           & 16.8          & 28.3          & 0.370          & 0.283          & 0.405           & 0.332          & 0.347       & 0.848          & 0.920          & 0.933           & \textbf{0.944} & 0.911          \\
\textbf{K = 512\phantom{0}}                                                                          & 30.0          & 50.5          & 47.8           & 29.7          & 39.5          & 0.386          & 0.300          & 0.411           & 0.339          & 0.359                & 0.837          & 0.917          & 0.932           & 0.942          & 0.907          \\
\textbf{K = 1024}                                                                         & \textbf{32.9} & \textbf{79.0} & \textbf{61.1}  & \textbf{43.5} & \textbf{54.1} & 0.560          & 0.355          & 0.417           & 0.363          & 0.423             & 0.823          & 0.895          & 0.930           & 0.936          & 0.896          \\ 
\cdashlinelr{1-16}
\textbf{soft-weighting}                                                                       & 20.6          & 30.4          & 33.8           & 8.7          & 23.3          & 0.346          & 0.267          & 0.397           & 0.337          & 0.336        & 0.857          & 0.922          & \textbf{0.935}  & 0.938          & 0.913          \\ \bottomrule
\end{tabular}
}
\label{tab:ablation_topk}
\end{table*}

\subsection{Soft-weighting}
\label{subsec:head_weighting}

This ablation study justifies the soft-weighting technique over selecting top-$K$ heads for ITI. The rationale for soft-weighting arises from the observation that the optimal set of top $K$ heads varies depending on the task or instrument. By adopting soft-weighting, we can bypass the need for hyper-parameter tuning specific to each task, simplifying the ITI process.

To validate this approach, we compare the performance of the full SMITIN (with soft-weighting) against SMITIN where the initial weights $w_{l,h}(t_0)$ are instead 1 for the top-$K$ most accurate heads according to probing (i.e., $(l,h)\in\mathcal{H}_K$), and 0 otherwise. Results from Table~\ref{tab:ablation_topk} indicate that while top-$K$ selection can yield good performance, it often requires task-specific tuning of $K$. In contrast, soft-weighting demonstrates a well-balanced performance across various metrics without necessitating such tuning. This balance is crucial because it means soft-weighting can adapt to various musical tasks and preferences, making ITI more flexible and user-friendly. 
However, it is important to acknowledge that soft-weighting may not always be the optimal choice for every ITI application. Users might prefer the top-$K$ approach depending on their specific needs and the nature of their musical goals.

\begin{table*}[t]
\centering
\sisetup{
    detect-weight,
    mode=text,
    tight-spacing=true,
    round-mode=places,
    round-precision=3,
    table-format=1.3,
    table-number-alignment=center
    }
\caption{
    Comparison with different intervention directions on \textit{Audio continuation}.
}
\setlength\tabcolsep{3.5pt}
\footnotesize
\begin{tabular}{
    l
    S[round-precision=1,table-format=2.1]
    S[round-precision=1,table-format=2.1]
    S[round-precision=1,table-format=2.1]
    S[round-precision=1,table-format=1.1]
    S[round-precision=1,table-format=2.1]
    c*{10}{S}
}
\toprule

& \multicolumn{5}{c}{\textbf{Success Rate [\%] ($\uparrow$)}}                                       & \multicolumn{5}{c}{\textbf{FAD$_\text{L-CLAP mus}$ ($\downarrow$)}}  &  \multicolumn{5}{c}{\textbf{Similarity ($\uparrow$)}}   \\ 
\cmidrule(lr){2-6}\cmidrule(lr){7-11}\cmidrule(lr){12-16}
\textbf{ITI Direction} & \textit{drums} & \textit{bass} & \textit{guitar} & \textit{piano} & \textit{avg.} & \textit{drums} & \textit{bass} & \textit{guitar} & \textit{piano} & \textit{avg.} & \textit{drums} & \textit{bass} & \textit{guitar} & \textit{piano} & \textit{avg.} \\ 
\midrule
\textbf{$\tilde{\theta}_{l,h}$} & 20.6 & 30.4 & 33.8 & 8.7 & 23.3 & 
0.346 & 0.267 & 0.397 & 0.337 & \textbf{0.336} & 
0.857 & 0.922 & 0.935 & 0.938 & \textbf{0.913} \\
\textbf{mass mean shift} & 17.0 & 48.6 & 36.0 & 9.7 & \textbf{27.8} &
0.355 & 0.291 & 0.398 & 0.323 & 0.341 &
0.855 & 0.923 & 0.928 & 0.941 & 0.911 \\
\bottomrule
\end{tabular}
\label{tab:obj_eval_com}
\end{table*}

\begin{table*}[t]
\centering
\sisetup{
    detect-weight,
    mode=text,
    tight-spacing=true,
    round-mode=places,
    round-precision=3,
    table-format=1.3,
    table-number-alignment=center
    }
\caption{
    FAD compared with the distribution of unconditioned generation on  \textit{audio continuation} and \textit{text-to-music}.
}
\setlength\tabcolsep{3.5pt}
\footnotesize
{
\begin{tabular}{
    l
    c*{10}{S}
    }
\toprule
& & \multicolumn{10}{c}{\textbf{FAD$_\text{L-CLAP mus}$($\downarrow$) w/. unconditioned}} \\
\cmidrule(lr){3-12}
\multicolumn{2}{c}{\textbf{Method}}
& \multicolumn{5}{c}{\textit{Audio continuation}}                                       & \multicolumn{5}{c}{\textit{Text-to-music}}   \\ 
\cmidrule(lr){1-2}\cmidrule(lr){3-7}\cmidrule(lr){8-12}
\textbf{ITI}                & \multicolumn{1}{l}{\textbf{Configuration}}                & \textit{drums} & \textit{bass} & \textit{guitar} & \textit{piano} & \textit{avg.} & \textit{drums} & \textit{bass} & \textit{guitar} & \textit{piano} & \textit{avg.} \\ 
\midrule
\multirow{1}{*}{\textit{\sffamily X}}
& \multicolumn{1}{l}{\textbf{\textit{``add \textless{}inst.\textgreater{}"}}} & 
0.042 & 0.026 & 0.026 & 0.064 & 0.03 &
0.013 & 0.009 & 0.011 & 0.009 & 0.01
\\ 
\cdashlinelr{1-12}
\multirow{3}{*}{\textbf{\checkmark}} & \multicolumn{1}{l}{\textbf{original ITI}} &
0.039 & 0.206 & 0.066 & 0.039 & 0.087 &
0.018 & 0.056 & 0.031 & 0.017 & 0.030
\\
& \multicolumn{1}{l}{\textbf{SMITIN ($\alpha=5.0$)}} &
0.022 & 0.017 & 0.018 & 0.010 & \textbf{0.016} &
0.012 & 0.009 & 0.010 & 0.006 & \textbf{0.009}
\\
& \multicolumn{1}{l}{\textbf{SMITIN ($\alpha=10.0$)}} &
0.079 & 0.042 & 0.033 & 0.014 & 0.042 &
0.045 & 0.017 & 0.019 & 0.008 & 0.022
\\ 
\bottomrule
\end{tabular}
}
\label{tab:obj_fad_uncond}
\end{table*}

\subsection{Intervention with Different Directions}
\label{subsec:appendix_center_of_mass}

Following \cite{iti}, we compare results using different intervention directions: logistic regression classifiers weights $\tilde{\theta}_{l,h}$ and mass mean shift. The intervention direction for mass mean shift is determined during probing as a vector that directs from the centroid of negative activations to the centroid of positive activations. The results are shown in Table~\ref{tab:obj_eval_com}. Similar to \cite{iti}, we observe that using mass mean shift as the intervention direction better achieves the target musical factor (according to success rate). However, we again observe a trade-off between controllability and generation distribution shift. Considering that generation quality is crucial in the music generation task, we adopt $\tilde{\theta}_{l,h}$ as the final intervention direction.

\subsection{FAD Compared with Unconditioned Generation}
\label{subsec:appendix_fad}

To support the claim that SMITIN does not significantly alter the original distribution of MusicGen, we compute the FAD score using the distribution of music outputs from unconditioned generation. For text-to-music, the comparison distribution is that of text-conditioned generation using MusicCaps text prompts, as discussed in the main paper. According to Table~\ref{tab:obj_fad_uncond}, the FAD score of \textit{``add \textless{}inst.\textgreater{}"} is significantly lower in \textit{text-to-music} than in \textit{audio continuation}, indicating that the output distribution between unconditional and conditional generation experiences a significant shift. We observe that SMITIN with $\alpha=5.0$ exhibits the closest distribution distance to the original MusicGen. This supports the claim that SMITIN achieves the desired controllability while retaining MusicGen's generation capability.

\end{document}